# A LOW-COST ISM-BAND MULTI-TRANSCEIVER COGNITIVE RADIO

A Thesis Presented

by

WILLIAM B. SCHOELER

Submitted in partial fulfillment of the requirements for the degree of

MASTER OF SCIENCE

Department of Electrical, Computer, and Systems Engineering

CASE WESTERN RESERVE UNIVERSITY

May 2022

# A LOW-COST ISM-BAND MULTI-TRANSCEIVER COGNITIVE RADIO

We hereby approve the thesis of

WILLIAM B. SCHOELER

candidate for the degree of Master of Science.

Committee Chair

Pan Li

Committee Member

Daniel Saab

Committee Member

An Wang

Date of Defense

April 7, 2022

*We also certify that written approval has been obtained for any proprietary material contained therein.

A Low-Cost ISM-Band Multi-Transceiver Cognitive Radio

ABSTRACT

by

WILLIAM B. SCHOELER

A Cognitive Radio is a type of Software-Defined Radio (SDR) that automatically detects available wireless spectrum and adjusts its physical hardware, modulation, or protocol parameters to obtain optimal throughput, latency, and range.  Much of prior Cognitive Radio research and design has required expensive transceivers using licensed bands that are not openly available for use by unlicensed users.  This thesis presents a low-cost hardware platform built from off-the-shelf components that utilizes free to use Industrial, Scientific, and Medical (ISM) bands, and implements a concurrent multi-spectrum point-to-point wireless protocol optimized for non-stationary devices.  Performance metrics such as cost, latency, throughput, and range are measured and analyzed.  Applications of such a wireless implementation are proposed and implemented, such as smart-city infrastructure that allows internet connectivity to inner-city users by providing Wi-Fi Access Points through mobile On-Board Unit (OBU) devices with uplinks delivered from stationary Roadside Unit (RSU) devices.

# TABLE OF CONTENTS

# LIST OF TABLES

# LIST OF FIGURES

# 1. BACKGROUND

## 1.1 ISM Frequency Spectrum Bands

Industrial, Scientific, and Medical (ISM) spectrum bands are ranges of frequencies that can be used by unregistered radiators in many locales around the world. ISM bands are uniquely identified by the fact that no royalty needs to be paid or license need to be acquired to utilize them. In comparison, non-ISM bands are licensed out by a regulatory agency to the highest bidder. In most countries, a regulatory agency exists that determines which frequency bands qualify as ISM bands and what rules apply to those bands. In the United States, devices that operate in these spectrum bands must abide by rules set out in FCC Part 15, such as maximum radiated power limits and required frequency hopping or digital modulation schemes.

Because ISM bands require no up-front regulatory approval (or a costly spectrum allotment from a regulatory agency), they are often used by low-cost or battery powered devices. Due to this ease of access, ISM band spectrum has become highly congested and contested due to the wide range of products and applications that operate in these bands. Wi-Fi, Bluetooth, Zigbee, LoRaWAN, Thread, as well as proprietary protocols occupy much of the commonly used 2.4 GHz ISM spectrum. In urban areas, congestion problems increase for static spectrum devices since there are more devices operating and vying for this shared spectrum in denser areas.

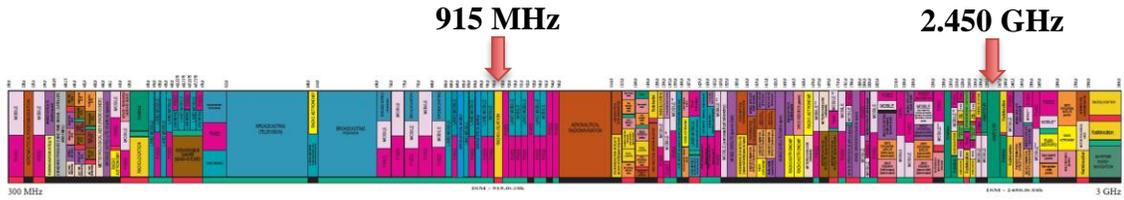

**Figure 1 – ISM Bands on FCC Allocation Chart**

In the United States, there are ISM Bands at 40 MHz, 433 MHz, 915 MHz, 2.4 GHz, 5.8 GHz:

**Table 1 - License Free ISM Bands Comparison**

| ISM Band | Technologies | Pros | Cons |
| --- | --- | --- | --- |
| 40 MHz | Not widespread consumer use, weather stations | Very long range | Low data rates, large antennas |
| 433 MHz | Not ISM in the United States, is in EU and Africa | Long range, less congested | Low data rates |
| 915 MHz | Z-Wave, TI "Sub 1GHz", Smart Home Automation | Less congested, good penetration | Mediocre data rates |
| 2.450 GHz | Bluetooth, Wi-Fi, Zigbee, Thread, LoRaWAN | Good data rates, many chipsets | Highly congested, poor penetration |
| 5.8 GHz | IEEE 802.11n Wi-Fi | Less congested, more bandwidth | Higher cost, lower range |

## 1.2 What is a Cognitive Radio?

A Cognitive Radio (CR) is a type of Software-Defined Radio (SDR) that can react to its environment by automatically adjusting its physical hardware, transmission modulation, or protocol parameters to provide optimal throughput, latency, and range in the spectrum available to the CR. A CR may measure its performance over time by using metrics that provide information about the quality of the connection to other radios, such as Bit Error Rate (BER), Frame Error Rate (FER), and Received Signal Strength Indicator (RSSI) to compute an overall Link Quality Indicator (LQI). CRs can fall under many subtypes, such as a Full Cognitive Radio (Mitola Radio) in which all parameters observed by a CR are considered, or a Spectrum-Sensing Cognitive Radio in which only the spectrum available to the CR is considered. In some CR networks, information and metrics are shared between nodes to determine the best way to communicate. Cognitive Radios are well suited to applications that require a high degree of connectivity and robustness, such as smart-city and campus wireless networks, emergency networks, medical applications, remote weather stations, and traffic control.

**1.3 ISM Usage in the Low-Cost Consumer Market**

Some research has posited that Cognitive Radios have not seen widespread consumer product adoption due to their high cost and complexity and use of non-ISM bands. As such, the most widely used low-cost ISM RF protocols are not Cognitive Radios. For example, although Wi-Fi does allow for setting many different channels, the ownership of this functionality is placed on the user and does not dynamically adjust based on the congestion of the surrounding frequency spectrum. Bluetooth does allow for channel hopping but does not initially sense the channels before deciding which channels to use in a channel hopping strategy. At present there is a consumer market hardware tradeoff in the hardware necessary to implement a Cognitive Radio–High-cost SDRs allow for maximum flexibility, but most low-cost ISM transceivers radios have far fewer features and may have limited throughput. Additionally, most high-cost SDRs require a host computer running Linux or comparable higher-order operating system, which further increases the cost of the system.

# 2. DESIGN

## 2.1 Goals

This thesis' goal is to design and implement a low-cost multi-transceiver ISM-Band Cognitive Radio that can automatically adjust its modulation and other protocol parameters to provide low latency, long-range, and reliable connectivity at greater than 1Mbps peak data rates on unlicensed ISM bands. The CR must be able to sense when a frame has not been successfully delivered and attempt to provide reliable delivery. The CR must have enough memory to maintain a list of frames in a buffer to re-route frames that have failed to be delivered within a certain time interval on its redundant transceivers as to provide maximum reliability of data transmission. Additionally, the CR must not interfere with other transmitters in the ISM domain, and as such must implement a CSMA/CA mechanism with binary exponential backoff. Lastly, the CR should be easy to interface to, allowing for a wide range of applications (no proprietary software should be needed to interface to the CR to send and receive arbitrary data). In summary, this thesis' resulting CR implements in a wireless link that is resilient to changes in the environment, and as such would be well suited to applications where either the one or both nodes in the wireless link are mobile and able to move throughout the environment.

## 2.2 Radio Protocol

By design, to achieve the lowest latency possible, an active connection that relies on time management and clock synchronization, such as Time Domain Multiple Access (TDMA) is not utilized. A major downside to the protocols leveraging TDMA is fact that the minimum latency in the system is related to the previously agreed-upon receiver wake up interval. For example, in Bluetooth spec this is an adjustable parameter from 7.5ms to 4s. While this can save power, as well as decrease on-air collisions in a star or mesh network topology, since this thesis' CR is optimized for point-to-point connectivity, TDMA is not necessary and adds extra latency overhead. Instead, the default idle state of any node in the protocol is receiving (RX) state. Unless a node is attempting to transmit a frame, or already receiving a frame, the node will remain in the RX state ready to receive a frame. Upon reception of a frame, the receiving node will verify the received frame's validity via a CRC check, and if valid, transmits an ACK for that received frame. All frames sent in the protocol have a data sequence number, and the ACK contains the corresponding frame's sequence number so that the transmitter knows which frames have been successfully received. This process is visualized in Figure 2.

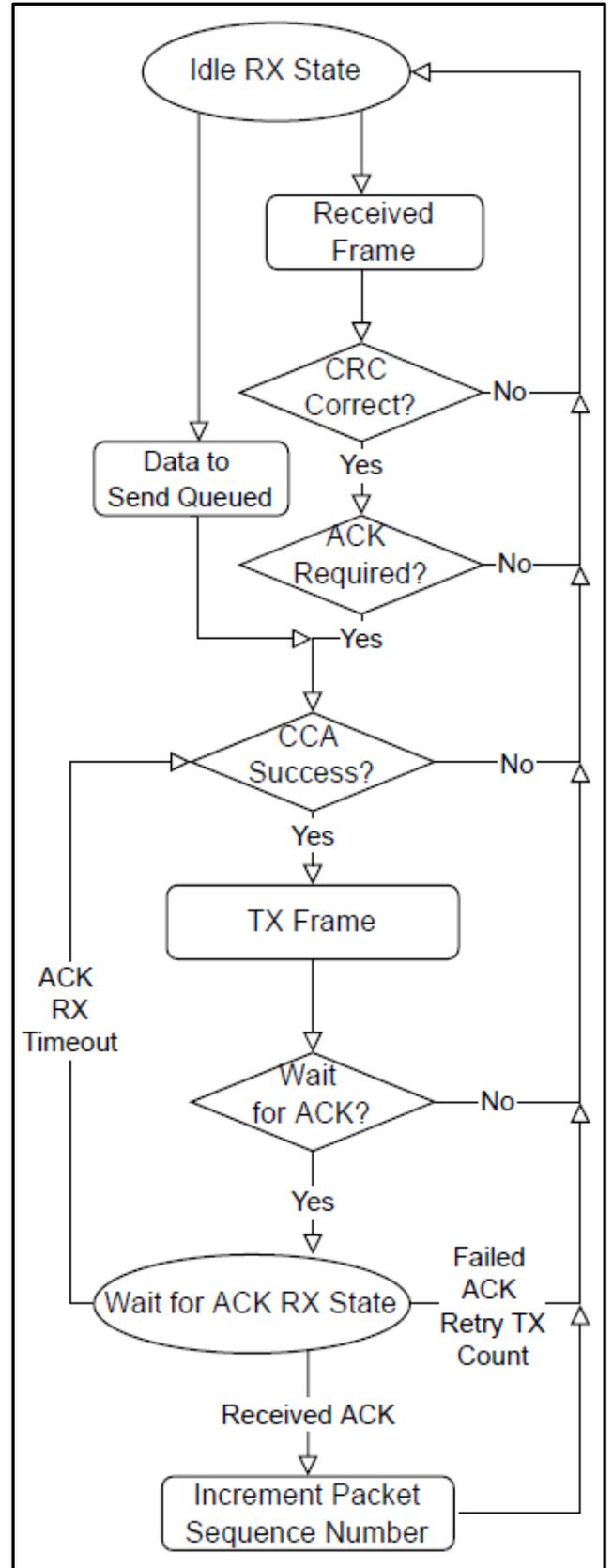

**Figure 2 – Protocol State Machine**

A key feature of the CR protocol is that if data is received on the CR's input queue while a receiving node begins to receive a frame, but before the CR has sent its corresponding ACK frame, the corresponding ACK frame will contain this newly inputted data to be sent. In this way, if there is a flurry of data being transferred from both nodes in the protocol at the same time, once the first successful frame has been received and acknowledged, both nodes can send frames with minimal delay. These transient connections during times of data throughput form an ad-hoc connection, where each participant of the protocol has a modulation-dependent and frame length dependent random backoff timeout which allows for non-congestive retry functionality.

When a frame is sent from one node to another and it is not acknowledged to the sender after some time, the frame is considered to have been lost and is called a NACK. If a NACK occurs, the node will attempt to frame resend again after a modulation-dependent binary exponential timeout. After a protocol-adjustable number of NACKs occur, a failed-to-be-sent frame is dropped entirely from the CR's buffer, and transmission is not attempted again. Because of this, if a certain CR user requires reliable delivery, a higher-level protocol with retry mechanisms like TCP must be used at the transport level. That being said, due to the fact that the CR has multiple transceivers, this dropped frame attempt to be transmitted on all transceivers available to the radio and may simply be delivered with a higher latency than normal when congestion occurs.

Because the CR has access to multiple transceivers simultaneously, it is also possible for more than one node to be transmitting or receiving frames in parallel. To allow frames received to be reordered in the original data stream order that the transmitting node received them in on the transmitters input queue, each frame contains metadata about the data stream order. One of these items is a sequence number, which is used by the receiving node to serialize the data stream back into the correct order. Frames received may not always be in the correct order by given by the time received, due to the fact that the frames may be received out of order because of frame retries due to interference. It is also possible that after a NACK timeout occurs that a frame may be retransmitted on a different transceiver, even if the original frame had been successfully received (but the ACK for this frame was lost). These duplicate frames must not be outputted on the data stream, and they are simply discarded.

The CR protocol is designed to be generic enough that it is replicable on various hardware. Although the implementation of the CR in this thesis contains two transceivers, more could be added to improve throughput and reliability. As such, any number of physical transceivers can be used in the Cognitive Radio, and the limit of how many transceivers would ultimately be bounded by the spectrum available to the user and interface bus speeds from transceiver to transceiver. Obviously, the transceivers chosen may or may not operate on the same frequency bands but must be used on different channels with adequate band gaps as to prevent any co-channel interference.

Various dynamic attributes of the CR protocol are adjustable based on each transceiver's link quality indicators. For instance, the length of frames being sent are dynamic, and based on the Frame Error Rate (FER) of recently sent frames. The basic idea behind this is that the shorter the length of a frame, the less time that is required for the transmitter to send its data, and thus there is less time for that frame to experience interference from the environment. As the FER increases, the frame length decreases, in an attempt to decrease FER.

Additionally, to increase the throughput of the CR, if the link quality is determined to be strong enough, a modulation shift will occur which results in a higher on-air data rate.  This hand off is a unique feature of the CR, as both nodes must switch over to the modulation simultaneously so that no frames are lost.  Modulation shifts can be initiated from either node in the CR but require an acknowledgement from the node being directed to change modulations before the modulation is initiated.

**2.2.1 Frame Structure**

The frame structure for the protocol is shown in Figure 3.  The first section of each frame contains a preamble, which enables the receiving node's PHY to identify the beginning of a frame being sent on-air.  After this is a sync word which must match a known 4-byte value.  Next is a 2-byte field that contains the number of bytes in the remainder of the frame, so that the PHY knows when a frame is completely received.  As such, the following field which contain the payloads has a dynamic length, up to a length of $2^{16}$.  Inside the payload is a 9-byte header that contains some metadata that describes the data the payload contains.  The remaining bytes in the payload contain the raw data that was received from the transmitters input queue.  Lastly, there is a 4-byte field which contains a 32-bit polynomial CRC over the contents of the entire frame, which the receiver can use to verify the validity of the received frame.

| Frame Fields | | | | | |
|---|---|---|---|---|---|
| Preamble 2 Bytes | Sync Word 4 Bytes | Payload Length 2 Bytes | Payload Header 9 Bytes | Payload Up to 1000 Bytes | CRC 4 Bytes |

**Figure 3 – Frame Fields**

The payload header field contains some control fields, such as a bit for resetting the sequence number (which is necessary when one node in the CR resets and the other does not), a bit for requesting a modulation upshift, and a bit acknowledging a modulation upshift request. 5 bits are reserved for future use out of the 8 bits in the control field. The other 8 bytes are for the frame sequence number, and for acknowledging received sequence numbers.

| Frame Payload Header Fields (9 Bytes) | | | | | |
|---|---|---|---|---|---|
| Sequence Reset 1 Bit | Modulation Shift Req 1 Bit | Modulation Shift Ack 1 Bit | Reserved for Future 5 Bits | Frame Sequence Number 4 Bytes | ACK Sequence Number 4 Bytes |

Figure 4 – Frame Payload Header Fields

**2.2.2 Frame Transmission Example Scenarios**

While there are many transmission scenarios in the protocol that the CR must be able to handle, the most common are below. In any RF protocol design, there is no guarantee that any frame will be sent successfully because at any moment a frame on-air may experience interference. As such, retries are included on CS failures, NACK failures, and NACK retry count timeouts. The CR increases in complexity compared to other protocols running on hardware in the same price range as our CR due to the fact that can be more than one transceiver active. Because of this, multiple frames can be "in-flight" simultaneously, and these must be arranged back in their original order, and duplicate frames must also be discarded.

**2.2.2.1 Node 0 TX to Node 1 RX with Successful ACK**

In this scenario, Node 0 receives data from its input queue to transmit immediately to Node 1. Node 0 is currently not receiving or transmitting any frames—it immediately attempts a Carrier Sense (CS) to detect if the spectrum is currently in use. In this scenario, the CS succeeds, and the frame is immediately transmitted. Node 1 is currently not transmitting or receiving any other frames, and successfully receives the preamble and frame contents over-the-air from Node 0. After a successful CRC check on the received frame, Node 1 sends this data to its output queue and transmits an ACK to Node 0. Node 0 successfully receives this ACK, and does not attempt retransmission:

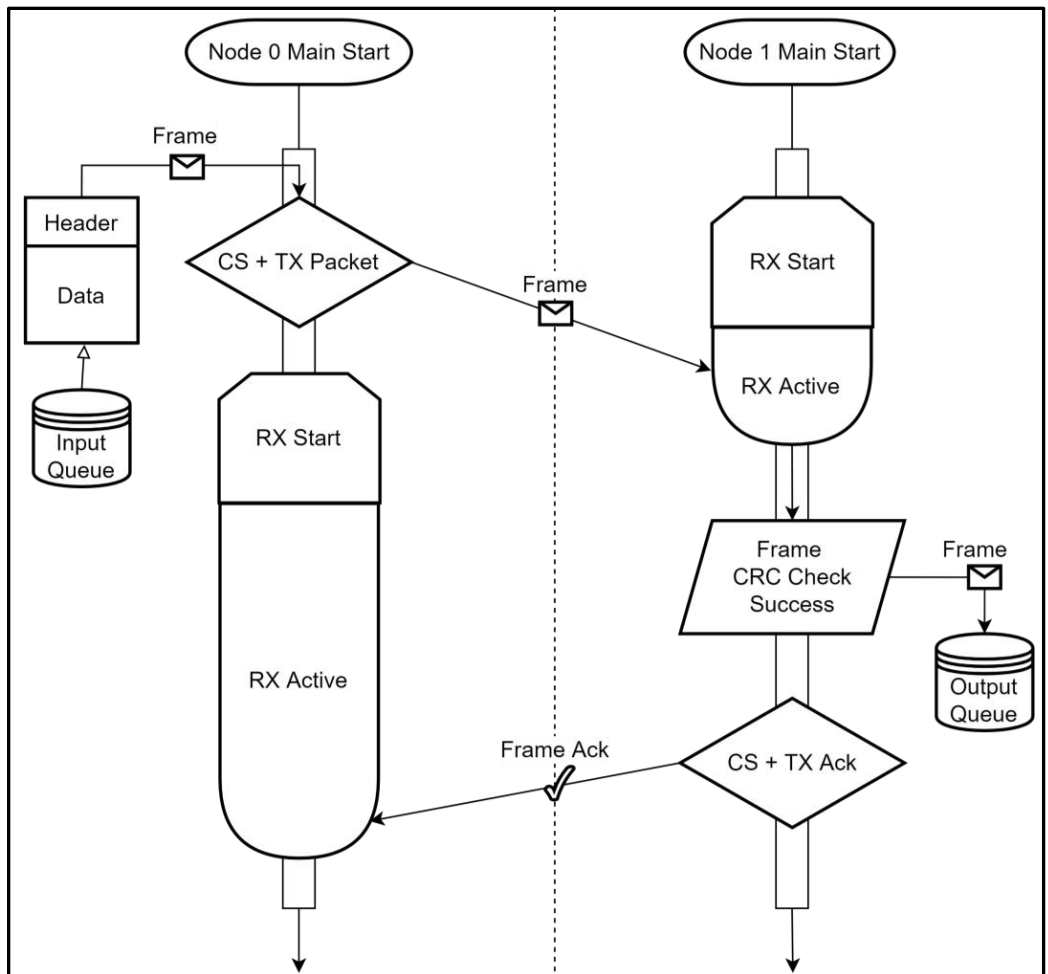

**Figure 5 – Node 0 TX to Node 1 RX with Successful ACK**

**2.2.2.2 Node 0 TX to Node 1 RX with Retry due to NACK**

In this scenario, the initial states of Node 0 and Node 1 are the same as scenario 1. However, the CS operation fails while the transmitter first checks to see if any other transmitters are currently active on the same channel. After a timeout, the frame transmission is retried and succeeds, after which Node 0 receives an ACK from Node 1.

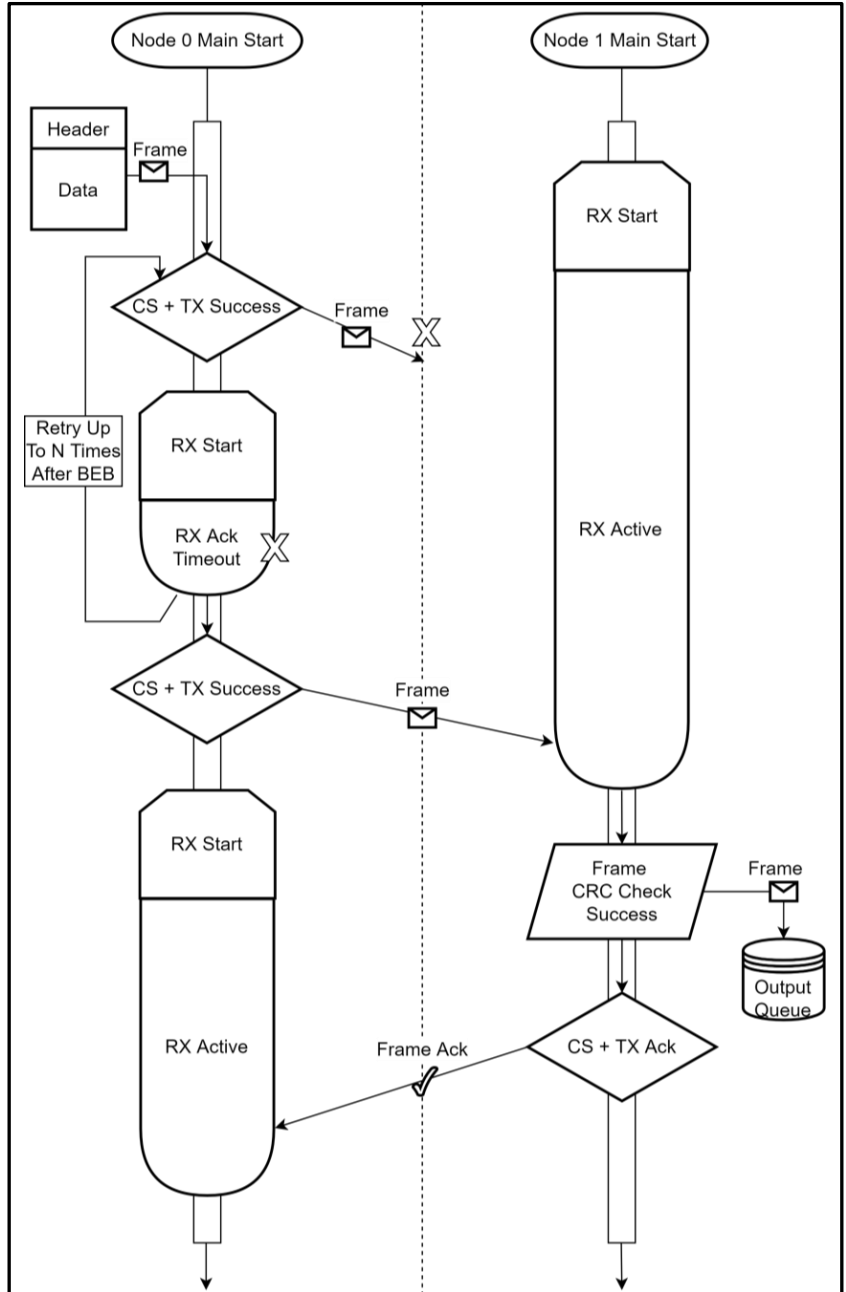

**Figure 6 – Node 0 TX with Retry due to NACK #1**

A slight variation of this scenario is presented in Figure 5. Instead of Node 1 failing to receive the frame from Node 0, Node 1 receives the frame successfully, but the ACK frame is intercepted and unable to be received by Node 0. As such, Node 1 attempts retransmission of Frame 1 until it receives a successful ACK.

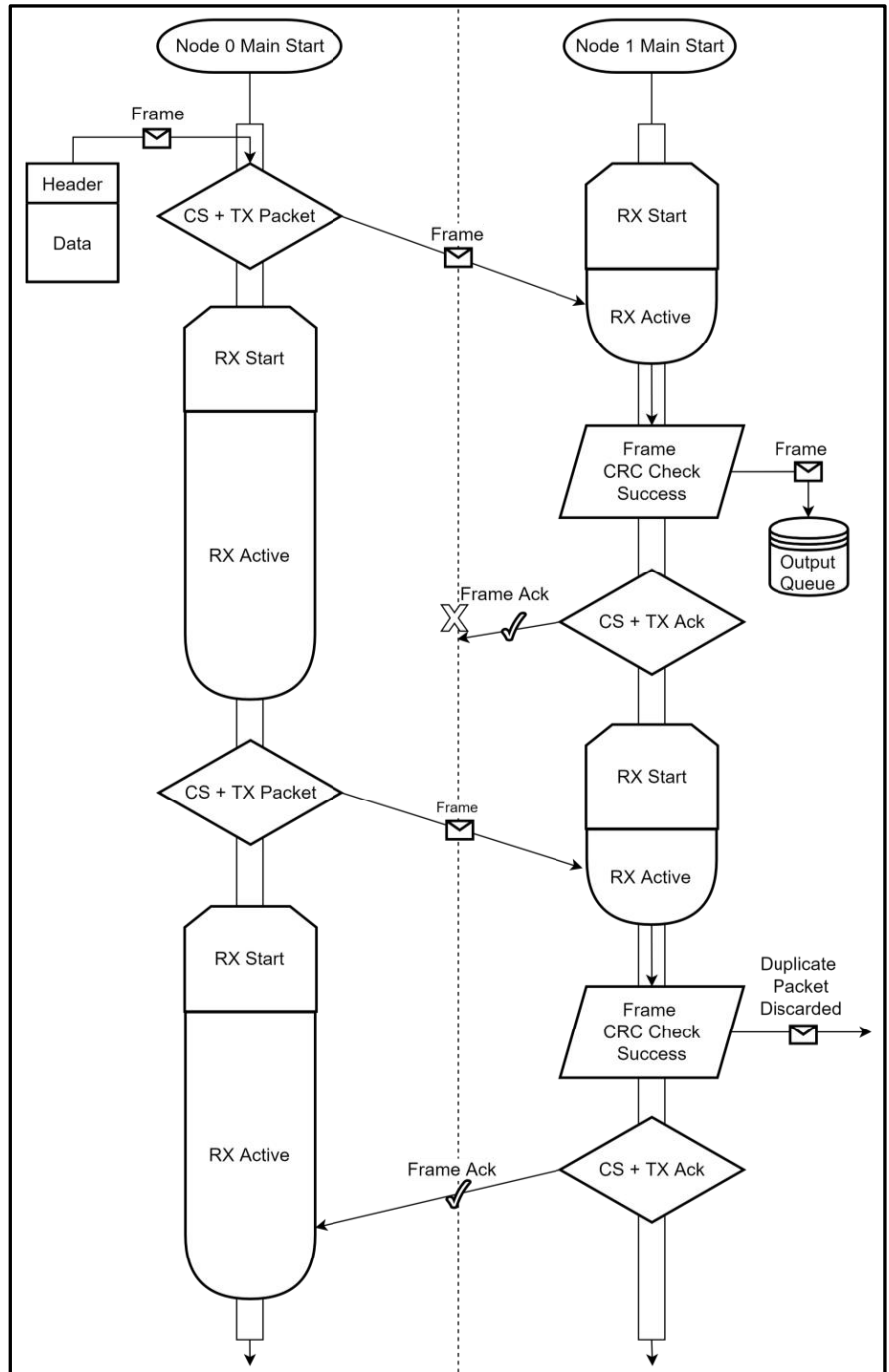

**Figure 7 - Node 0 TX with Retry due to NACK #2**

**2.2.2.3 Node 0 TX to Node 1 RX/TX to Node 0 RX with ACK**

Sometimes, the transceiver simultaneously receives data from its input queue while it is already receiving data. When this occurs, instead of the node sending an ACK and this new data in two separate frames, the node transmits both items in one frame.

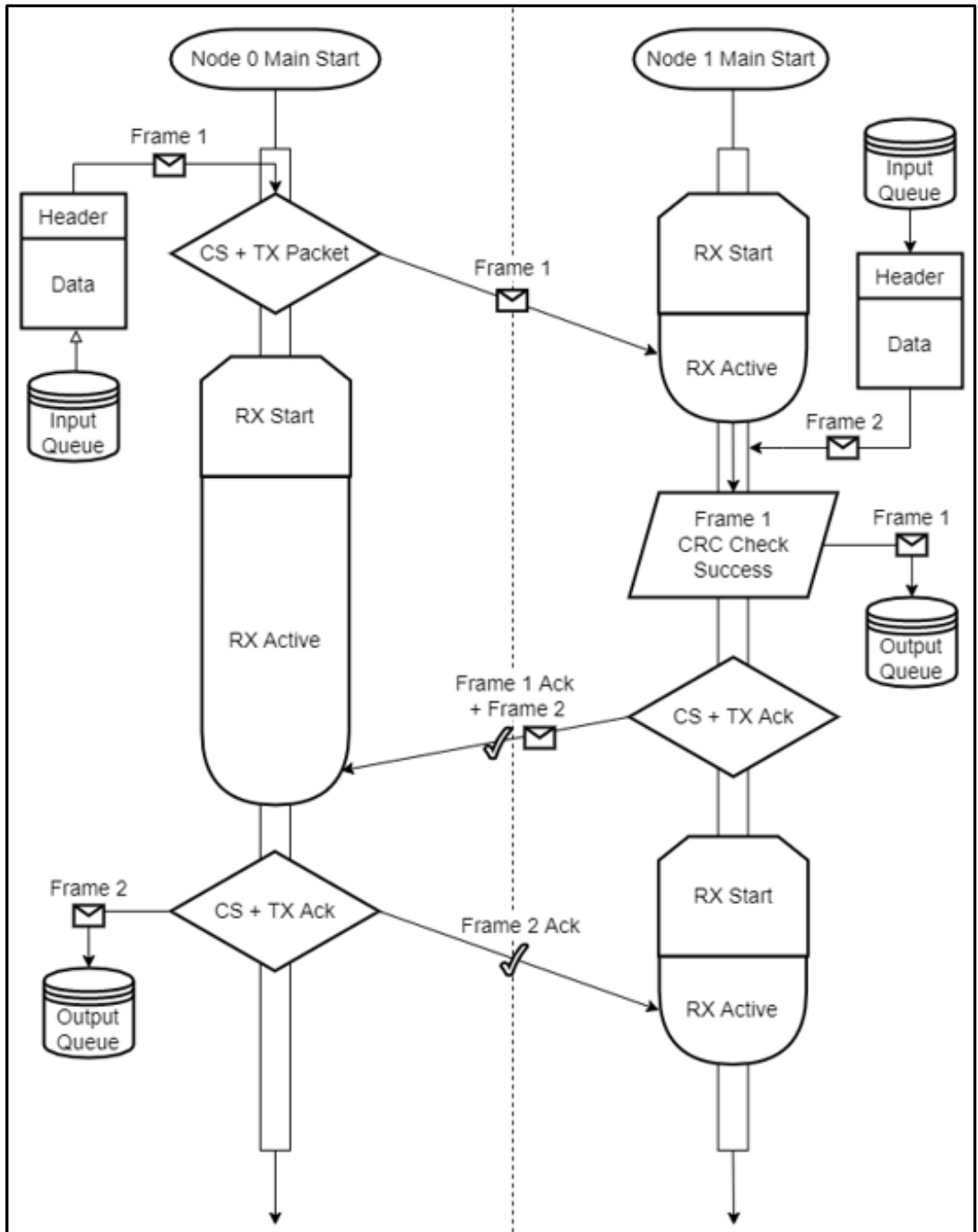

**Figure 8 - Node 0 TX to Node 1 RX/TX to Node 0 RX with ACK**

## 2.2.2.4 Node 0 TX with Retry on Node 0 Helper due to TX Retry Count Failure

When continual interference occurs on the channel of one transceiver, the frame will eventually fail and be dropped. When this occurs, the frame will be routed to the secondary transceiver, which will then attempt to transmitter the frame on a different channel.

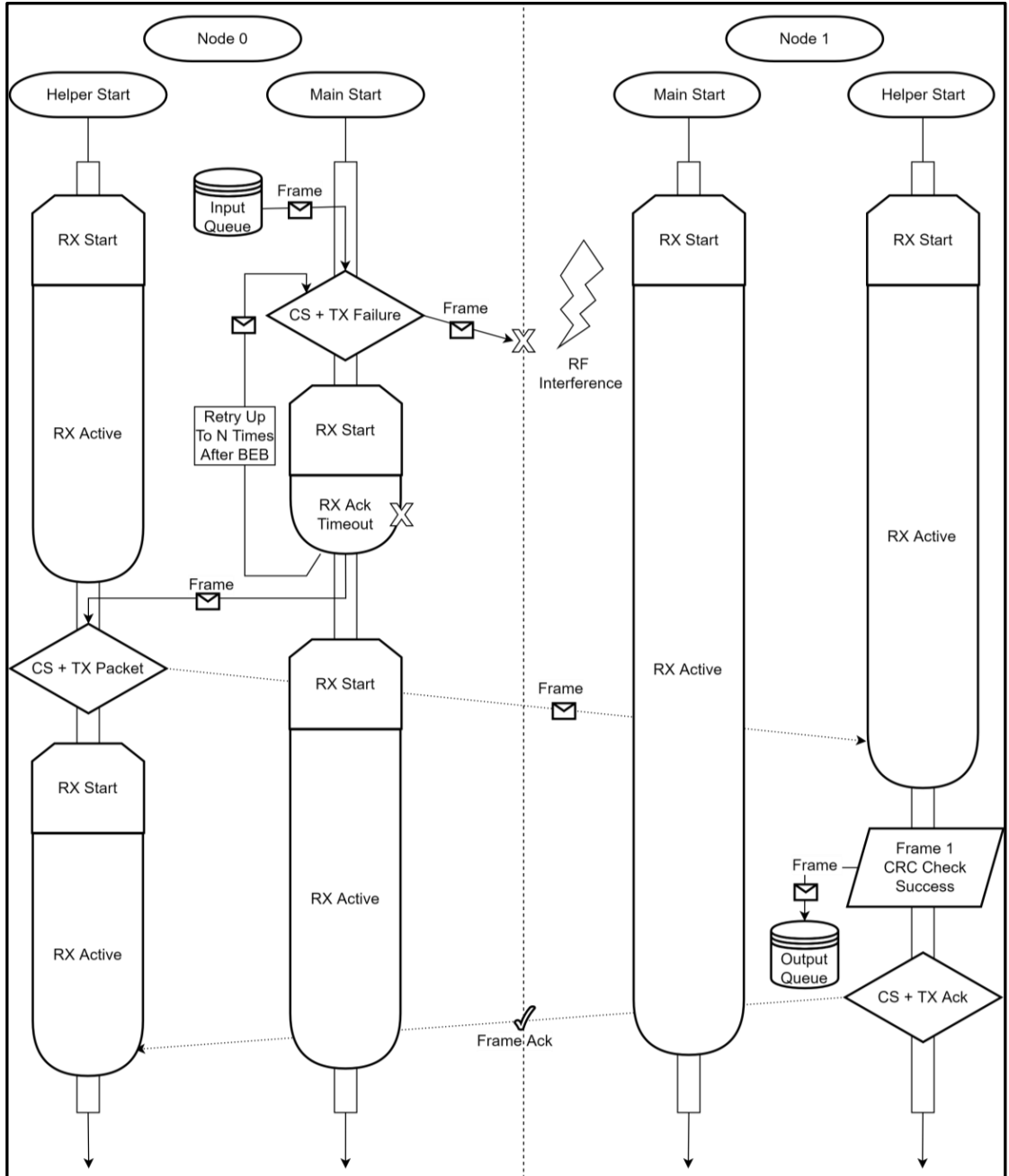

**Figure 9 - Node 0 TX with Retry on Node 0 Helper due to TX Retry Count Failure**

### 2.2.3 Dynamic frame lengths

Frames of fixed length are not spectrally efficient, because a fixed length frame could require a frame to be larger than the actual data it contains. Furthermore, large frames are not ideal in congested environments because they will have longer on-air-transmission times. $t_{OA}$, or the On-Air Time, is related to a frame's length ($l$) in bytes, and the transmitter's data rate ($d$) in bits per second. Data rate is usually a constant and determined by a transceiver's modulation.

$$t_{OA} = l \times \frac{d}{8}$$

The likelihood that a frame will experience interference increases with the frame's on-air transmission time, since at any moment another user in the ISM band may attempt to transmit at the same time. Further compounding this problem is the fact that not all devices in the ISM bands have a requirement to perform a Clear Channel Assessment (CCA) before transmitting. Additionally, due to the hidden node problem, a device that is CCA capable may not know that the CR is transmitting and may inadvertently jam the CR receiver. For these reasons, variable length frames are necessary. The CR transmitter prepends length information to each frame that is transmitted so that the CR receiver is aware of how long to listen. Before sending a frame, any node in the system can change a frame's length depending on link quality indicators.

The CR implements simple logic to adjust the next frame's length dynamically. If an ACK is received, the next frame's length will be increased by 10%, up to a maximum frame length which is dependent on the current transceivers modulation data rate (and thus, $t_{OA}$). If a NACK is received, the next frame's length will be decreased by 10%, to a minimum of 250 bytes.

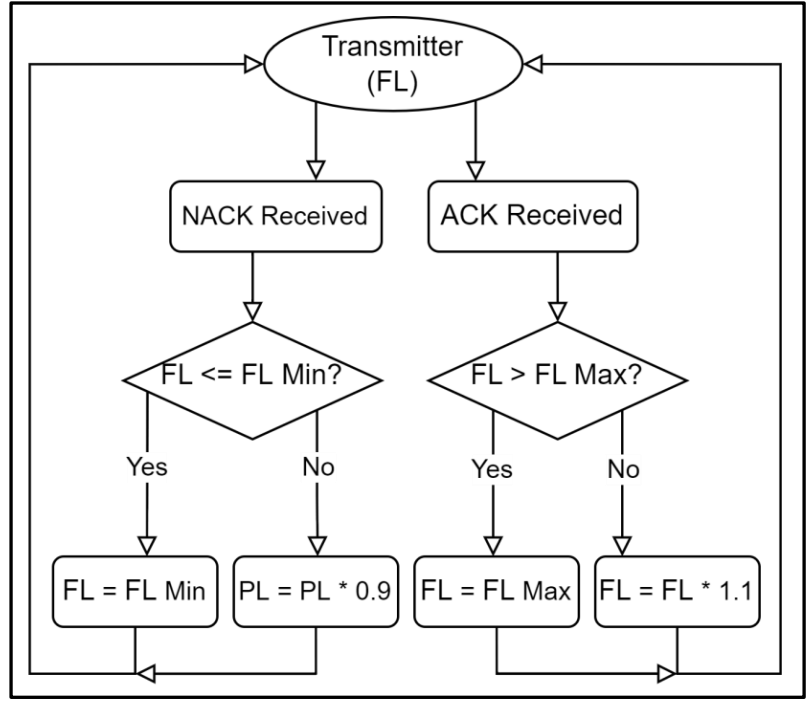

Figure 10 – Dynamic Max Frame Length Flowchart

| Modulation Data | Max Frame Length | On-Air Time ($t_{OA}$) |
|---|---|---|
| 915 MHz 200 Kbps | 250 Bytes | 10 ms |
| 915 MHz 1 Mbps | 1000 Bytes | 8 ms |
| 2.4 GHz 1 Mbps | 1000 Bytes | 8 ms |
| 2.4 GHz 2 Mbps | 1000 Bytes | 4 ms |

Table 2 – Maximum Frame Length and On-Air Time per Modulation

**2.2.4 Link Quality Indicators & Modulation Shifting Logic**

The CR leverages Link Quality Indicators (LQI) to determine when to change modulation schemes. Some examples of Link Quality Indicators are Bit Error Rate (BER), Frame Error Rate (FER), and Received Signal Strength Indicator (RSSI). The CR uses a combination of FER and RSSI to make the decision of when to go up or down in a modulation scheme, and when to increase or decrease the frame length.

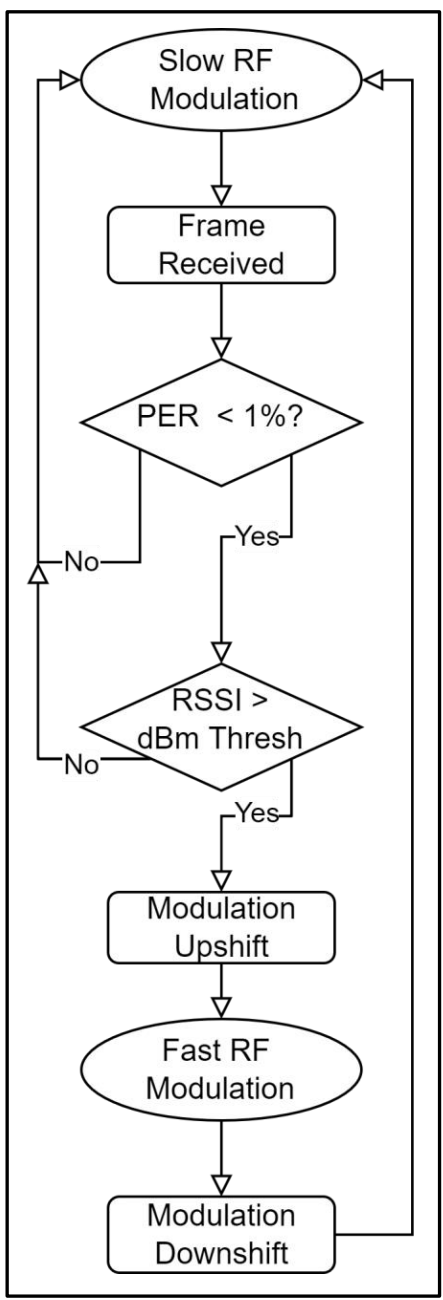

**Figure 11 – Modulation Shift Logic Flowchart**

**Table 3 – Modulation Increase Logic Table**

| Modulation | Sensitivity (dBm) | # Of Back-to-Back Successful Frames |
|---|---|---|
| 915 MHz 200 Kbps | -100 | 10 |
| 915 MHz 1 Mbps | -93 | N/A |
| 2.4 GHz 1 Mbps | -94 | 20 |
| 2.4 GHz 2 Mbps | -89 | N/A |

# 3. IMPLEMENTATION

## 3.1 Hardware

After extensive market research, the Texas Instruments (TI) CC1352 was chosen to implement the design due to its low cost, ability to operate on both the 915 MHz and 2.4 GHz ISM Bands, and excellent receiver sensitivity. This transceiver provides a small microprocessor with 80KB of RAM to implement the desired protocol logic and to buffer the data stream inputs and outputs. Most critically, it has a separate radio processor that allows the MAC and PHY to run independently of the protocol logic microcontroller which allows the MAC and PHY to meet the tight timing requirements required for the protocol.

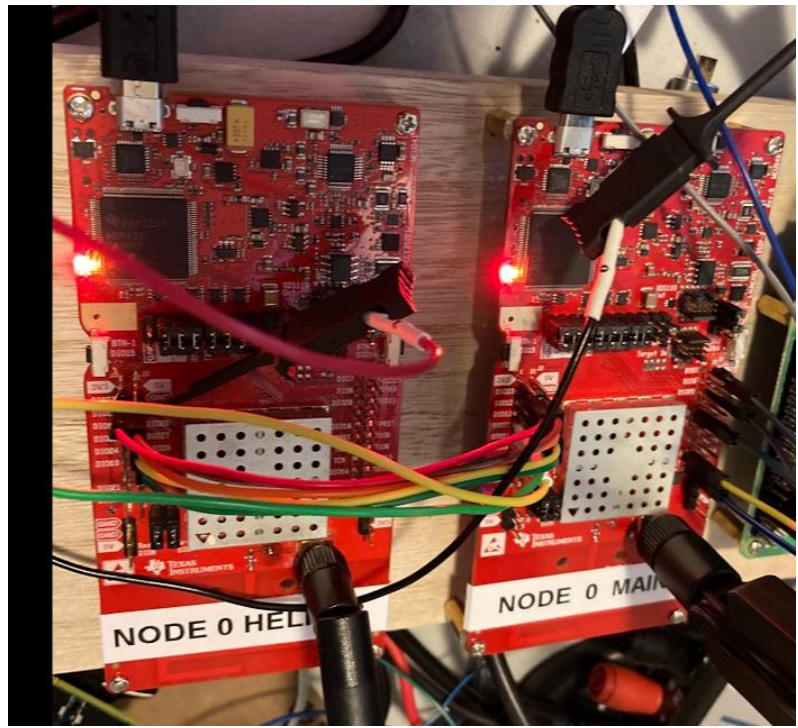

**Figure 12 – Cognitive Radio Node 0, 2x TI CC1532P**

The implemented CR contains two discrete hardware devices, the "Main" device, and the "Helper" device; together they will be referred to as a "node". Each device contains a physical transceiver operating on non-overlapping channels with an appropriate guard band. In this implementation, the Main operates at 2440 MHz with 1 MHz bandwidth and utilizes GFSK modulation for both the 1Mbps and 2Mbps data rates. The Helper operates at exactly 915 MHz and utilizes GFSK with a 271.3 kHz Bandwidth at 200Kbps data rates and 2185.1 kHz of Bandwidth at the 1Mbps data rate. The Main contains the software task that implements radio protocol logic described in section 2.2, The Main also contains the software task that parallelizes the serial input data stream and another task that reserializes the received on-air data from both itself and the Helper. The

The Main and the Helper are interfaced to each other via a secondary UART interface with Hardware Flow Control, running at 2Mbps. Hardware flow control is necessary since the Helper device may no longer be able to receive any more data if there is interference, and the Main must not send more data to the Helper during this time, or the data will be lost. The communication interface pinout connections are docuemented in Table 3 and Table 4.

**Table 4 – Main Device Pinout**

| Modulation | Pin | Color |
|---|---|---|
| Host Comm UART CTS | DIO16 | Purple |
| Host Comm UART RTS | DIO11 | Blue |
| Host Comm UART TX | DIO18 | Green |
| Host Comm UART RX | DIO17 | Yellow |
| RF Comm UART CTS | DIO23 | Purple |
| RF Comm UART RTS | DIO27 | Blue |
| RF Comm UART TX | DIO25 | Green |
| RF Comm UART RX | DIO26 | Yellow |

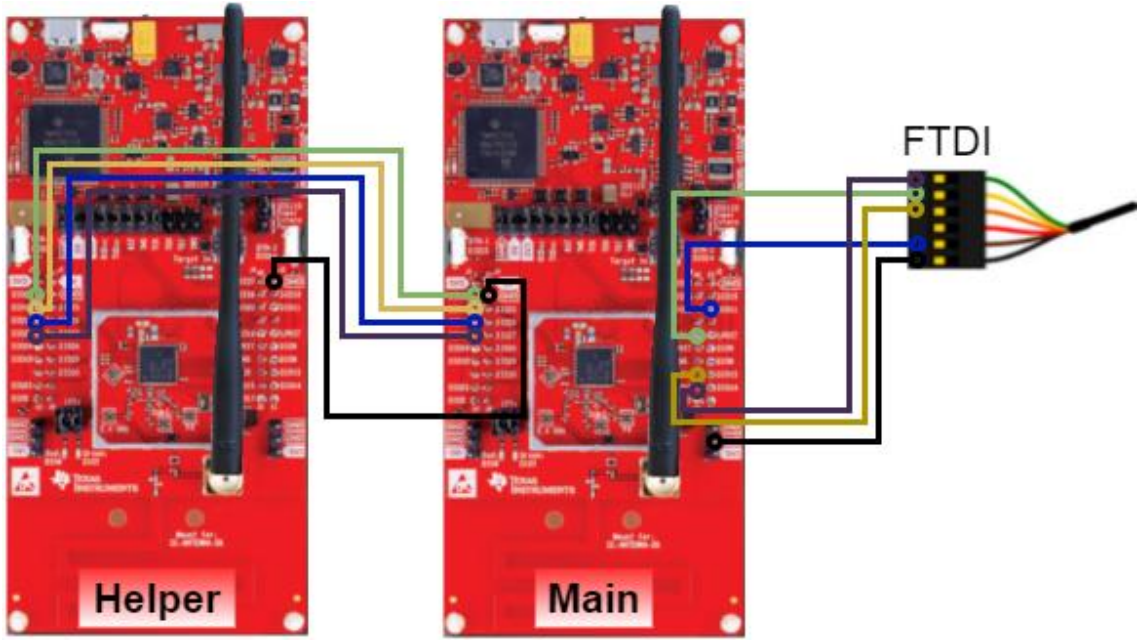

**Figure 13 – Main and Helper Device Pin Connections**

**Table 5 – Helper Device Pinout**

| Modulation | Pin | Color |
|---|---|---|
| RF Comm UART CTS | DIO27 | Blue |
| RF Comm UART RTS | DIO23 | Purple |
| RF Comm UART TX | DIO26 | Yellow |
| RF Comm UART RX | DIO25 | Green |
| Modulation | Pin | Color |
| RF Comm UART CTS | DIO27 | Blue |
| RF Comm UART RTS | DIO23 | Purple |
| RF Comm UART TX | DIO26 | Yellow |
| RF Comm UART RX | DIO25 | Green |

The implemented CR accepts input and output serial data over a bi-directional Universal Synchronous Asynchronous Receiver Transmitter (UART) running at a clock speed of 2 Mbps. To prevent loss of data, the UART leverages hardware flow control so that data will not be accepted via the input if there is no room in the radio's input buffer. This can happen at times of high interference or high data rates. Once buffered data has been successfully sent to the intended receiver, and an acknowledgement has been received by the sender, free buffer becomes available and the UART indicates to the host device that it is able to accept new data to be transmitted. After a certain number of failures to transmit, data in the input buffer is dropped to allow retrying transmissions with new data, as transport layer protocols normally implement their own retry mechanisms.

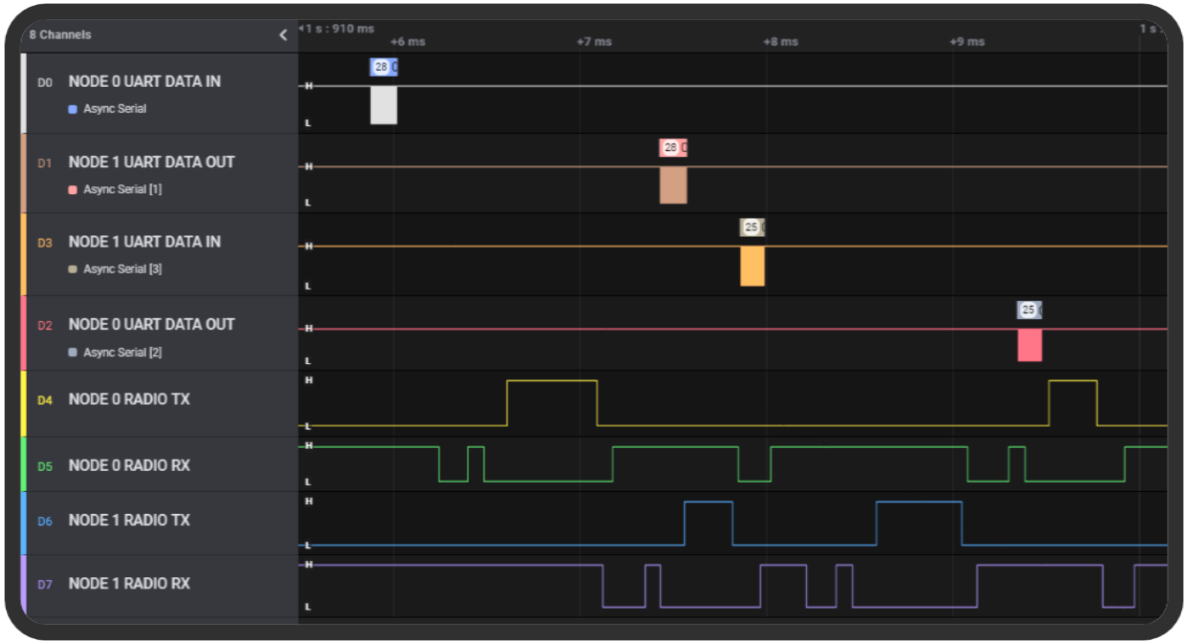

**Figure 14 – Asynchronous TX from Node 0 to Node 1 Followed by Node 1 to Node 0**

A Clear Channel Assessment (CCA) is implemented using Carrier Sense (CS) radio functionality to prevent interference with other devices operating in the ISM bands the CR transceivers occupy. Before transmission of any frame, a CS is performed to determine that the channel about to be utilized by the transmitter is free to use and has a lower probability of being interfered or jammed.

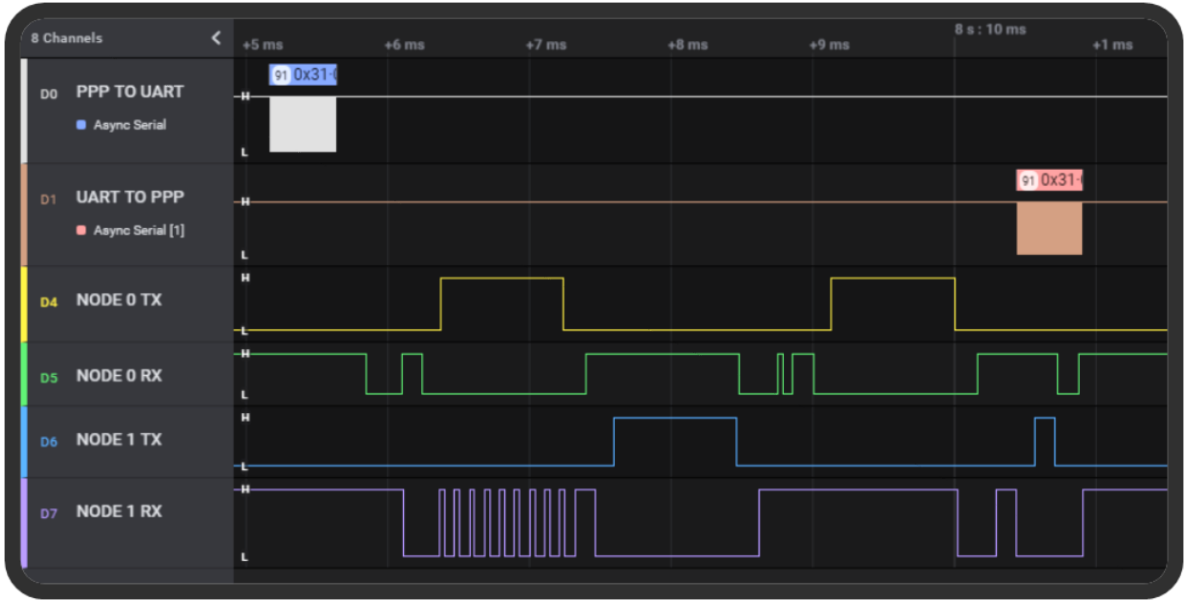

**Figure 15 – Node 1 CS Failure Until Node 0 TX Completion**

To minimize protocol processing time, a proprietary TI command chaining feature of the CC1352 chipset is leveraged to minimize processing latency. This offloads certain radio logic decisions from the MCU to an internal radio processor. For example, since before every transmission a CS check must be performed, the CS command is chained to the transmission (TX) command. The command chaining logic engine allows for a failure of the CS to prevent a TX after a certain amount of time. On the other hand, if the CS is successful, the TX begins immediately without any further indication or processing needed from the MCU. Command chaining, or some other mechanism where the logic on when to begin or not begin a transmission resides in hardware or other high-speed peripherals is critical to implement tight timings with low jitter that the protocol requires.

**3.2 Software Architecture**

Due to the tight timing requirements necessary to implement the presented CR protocol, a Real-Time Operating System (RTOS) is utilized rather than a higher order operating system such as Unix. The TI CC1352P provides several options for RTOS choices, FreeRTOS or TI-RTOS. TI-RTOS was ultimately selected since Texas Instruments provides useful drivers for various hardware peripherals with this OS, such as the UART interface and command chaining to the hardware of the Radio's RF Core.

An RTOS allows for more than one logical piece of code (called a "task") to execute as though it were the only program running sequentially on the system. The application programmer can then write multiple tasks that interact with each other through a network of queues and semaphores and allow the RTOS to parallelize them automatically. However, there is still only one logical CPU core on the CC1352P, which means only one task can be running at any given time. As such, the implemented CR leverages the Priority Preemption feature of the RTOS which allows the programmer assign arbitrary execution priorities to each task, which informs the RTOS of which tasks should run first if there is more than one task ready to run. In general, the task with the tightest timing requirements should have the highest task priority in the system, as it will run before the others.

### 3.2.1 Tasks

The Main device contains six tasks running simultaneously. One of the tasks, RF Control, implements the CR protocol described in section 2.2. RF Control accepts data from the Parallelizer task. The Parallelizer determines which physical transceiver any given frame should be routed to. Additionally, the RF Control task may receive asynchronously receive a frame from the other node and send this data the Sequencer. The Sequencer holds on to received frames for up to 300ms and attempts to sequence the frames outputted to the host device in order according to received frame's sequence number. If a sequence number is missing in a sequence of frames after no new frames have been received for 300ms, or the sequencer's buffer becomes full, all received frames are dumped to the host device. The other three tasks, UART Data Input, UART RF Helper Comm Input, and UART RF Helper Comm Output arbitrate the data handing between the Helper and the input data queue from the host device.

**Table 6 – Main Device Tasks**

| Task Name | Priority | Description |
| --- | --- | --- |
| RF Control | 6 | Implements Radio Protocol from section 2.2 |
| Sequencer | 5 | Sequences received frames back order to host |
| Parallelizer | 4 | Routes frames to transmitters (Main or Helper) |
| RF Helper UART Input | 3 | Reads from RF Helper UART Interface |
| RF Helper UART Output | 2 | Writes to RF Helper UART Interface |
| Host UART Input | 1 | Reads from Host UART Interface |

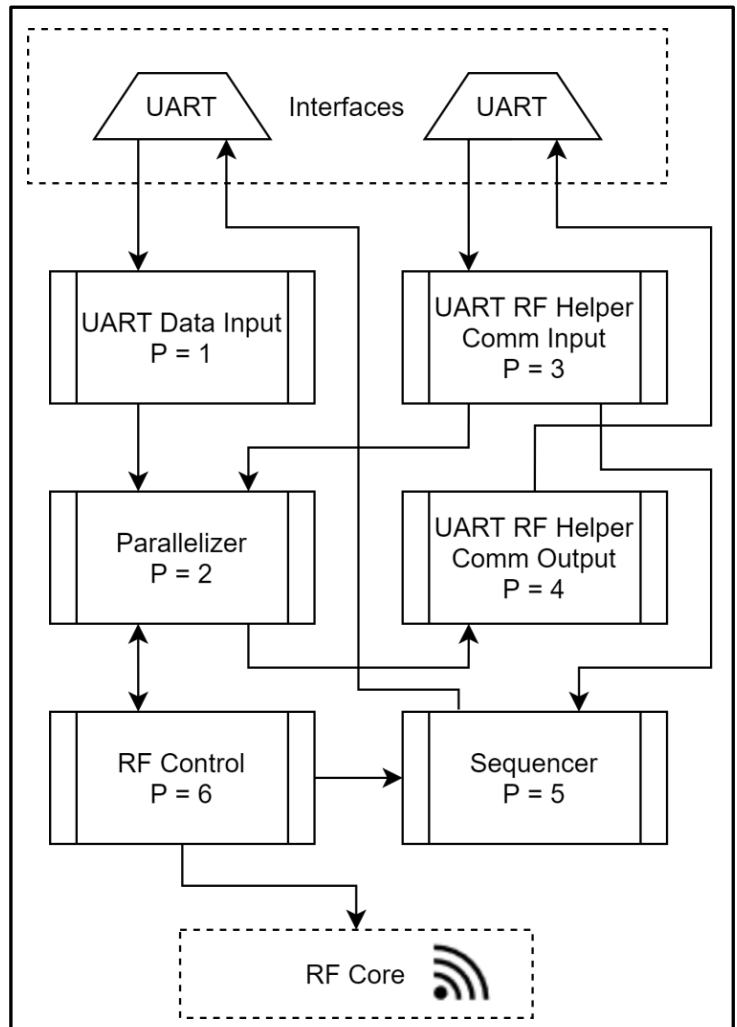
**Figure 16 – Main Device Task Diagram**

The Helper device has three tasks. The RF Control is an exact copy of the task on the Main device, but instead of interfacing to the Sequencer and Parallelizer directly on device, the Helper sends and receives frames to the Sequencer and Parallelizer to the Main over the secondary UART interface. The other two tasks handle receiving data from the Main device and sending data to the Main device.

**Table 7 – Helper Device Tasks**

| Task Name | Priority | Description |
|---|---|---|
| RF Control | 6 | Implements Radio Protocol from section 2.2 |
| RF UART Input | 3 | Receives frames from Main |
| RF UART Output | 4 | Routes frames to Main |

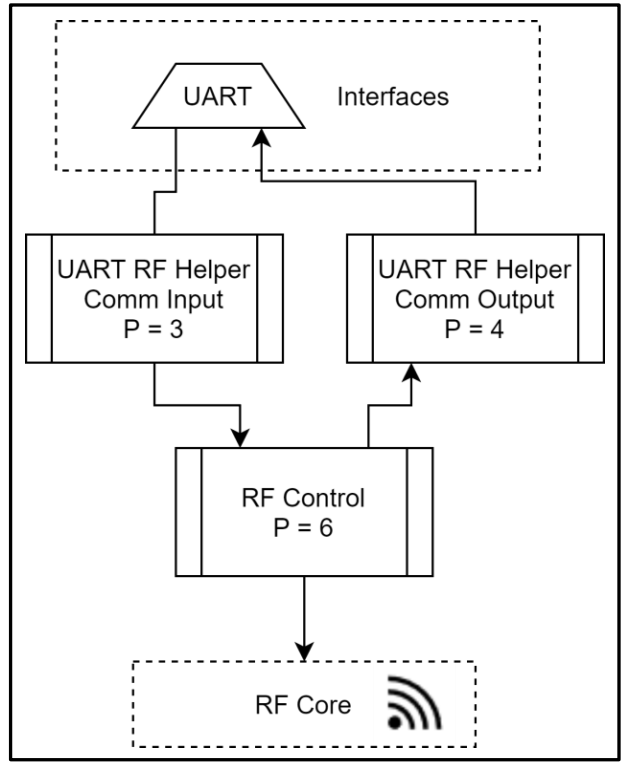

**Figure 17 – Helper Device Task Diagram**

**3.2.2 Dataflow**

A complete view of the Main and Helper (a "node") from a software perspective is given in Figure 14. The Main device accepts input data and can output data to the host over a UART interface, and the Main and Helper are connected via another UART interface.

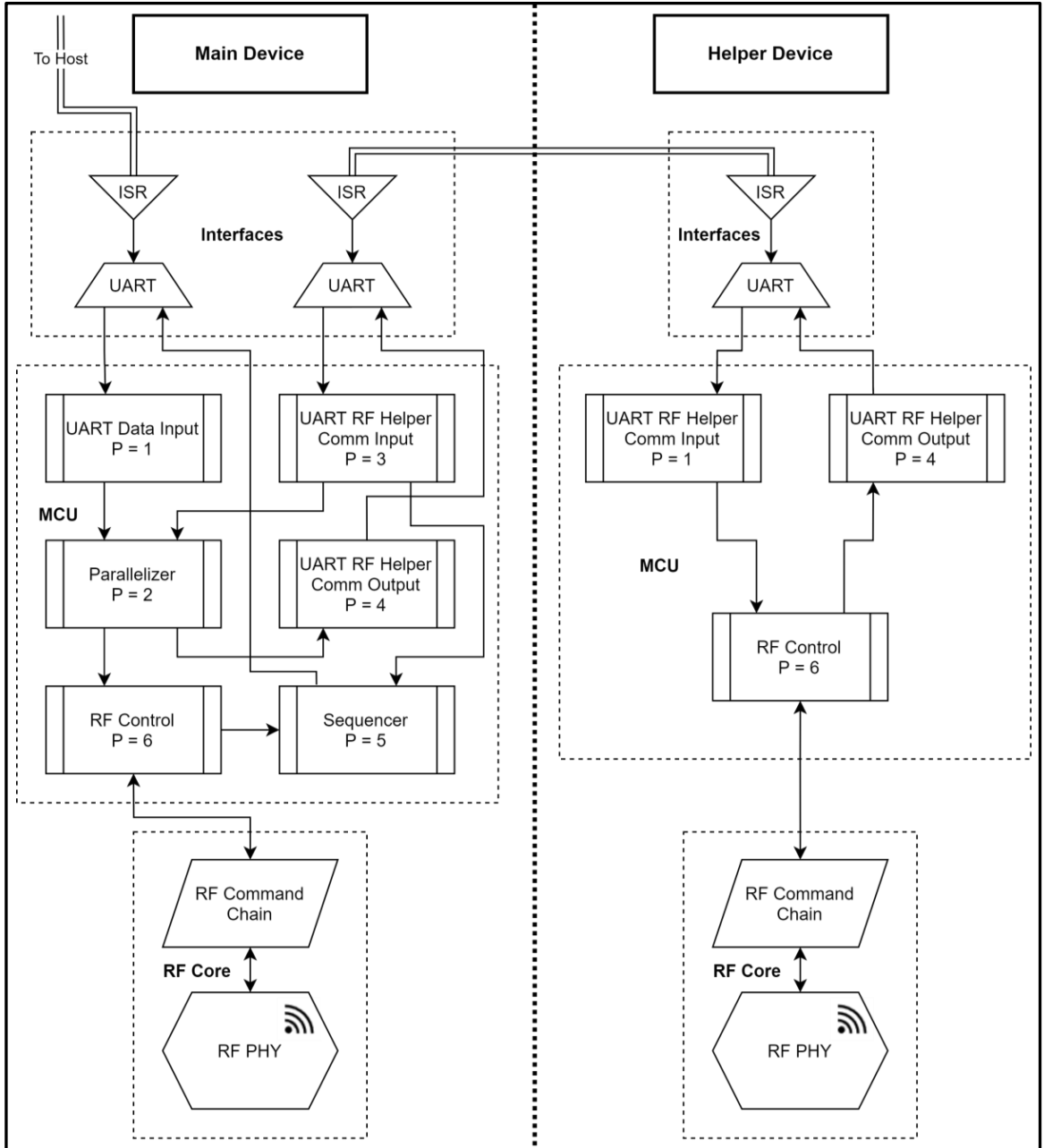

**Figure 18 – CR Tasking and Interrupt Diagram with Interfaces**

# 4. ANALYSIS

## 4.1 Testing Infrastructure

A test setup was required to be constructed to properly test and characterize the performance of the implemented CR design. To replicate consistent testing conditions, two UART to FTDI interfaces were utilized. The FTDI interfaces connect to a host PC. Our test fixture connects one FTDI cable to one node, and another FTDI cable to the other node. Three main test suites were performed to verify the implemented CR design:

1) Sending random data from Node 0 to Node 1, verifying that the data received on Node 1 is identical to the data sent from Node 0
2) Sending and receiving random data from Node 0 to Node 1 and vice versa simultaneously
3) Sending and receiving a previously sniffed Point-to-Point Protocol (PPP) data transfer that the prototype in section 5.1 leverages to pass TCP/IP frames to and from the internet

Using scripts implemented in python, random data is sent from the host PC to both nodes simultaneously. The script on the host PC then read the FTDI and verifies that the data received matches the data sent. With a good connection in a non-noisy environment, these tests have been verified to run hours at a time with no errors.

## 4.2 Characterizing the Implementation

### 4.2.1 Maximum Theoretical Throughput Given Hardware Specific Transceiver Delay

Although the CR protocol is designed to be hardware agonistic, there are some hardware-specific parameters that must be measured. Any transceiver contains inherent delays when switching from a RX state to a CS state, from the CS state to the TX state, or from a TX state back to an RX state. Additionally, since the CR can change modulations dynamically, there is delay when switching from RX to a new modulation called Frequency Synthesis (FS).

| Time Delay | 915 MHz PHY | | 2.4 GHz PHY | |
|---|---|---|---|---|
| | 200 Kbps | 1 Mbps | 1 Mbps | 2 Mbps |
| $t_{RXend} \to t_{CSstart}$ | 240.7 µs | 231.8 µs | 234.1 µs | 459.2 µs |
| $t_{CSstart} \to t_{TXstart}$ | 213.1 µs | 214.7 µs | 212.4 µs | 213.8 µs |
| $t_{TXstart} \to t_{TXend}$ (full pkt) | 40.93 ms | 8.11 ms | 8.12 ms | 4.08 ms |
| $t_{TXend} \to t_{RXstart}$ | 80.7 µs | 82.9 µs | 82.0 µs | 80.5 µs |
| $t_{RXstart} \to t_{RXend}$ ($ack\ only$) | 2.61 ms | 505.6 µs | 506.6 µs | 696.8 µs |
| $t_{RXend} \to t_{FS} \to t_{RXstart}$ | 392.1 µs | 562.5 µs | 633.1 µs | 581.3 µs |

**Table 8 - Hardware Specific Transceiver Delays**

These delays reduce the effective data throughput of the CR and can be used to find the maximum theoretical throughput of the data link by using the following equations, where 8000 represents the number of bits in a completely full frame:

$Throughput\ (bits/s, unidirectional) = 8000\ /\ ((t_{RXend} \to t_{CSstart}) +$
$(t_{CSstart} \to t_{TXstart}) + (t_{TXstart} \to t_{TXend}) + (t_{TXend} \to t_{RXstart}) + (t_{RXstart} \to t_{RXend}))$

$$Throughput\ (bits/s, bidirectional) = 8000\ /\ ((t_{RXend} \to t_{CSstart}) +$$

$$(t_{CSstart} \to t_{TXstart}) + 2(t_{TXstart} \to t_{TXend}) + (t_{TXend} \to t_{RXstart}))$$

For the bidirectional case, $(t_{RXstart} \to t_{RXend})$ is replaced with $(t_{TXstart} \to t_{TXend})$ since this time will be equal to a full-length frame plus an ack.

By using the data collected and formulas given above, we can solve for the maximum theoretical throughput of the two transceivers in the implemented CR:

$Throughput\ (bits, unidirectional, 915\ MHz\ @\ 200\ Kbps) = 181{,}510$ bps

$Throughput\ (bits, bidirectional, 915\ MHz\ @\ 200\ Kbps) = 97{,}093$ bps

$Throughput\ (bits, unidirectional, 915\ MHz\ @\ 1\ Mbps) = 874{,}794$ bps

$Throughput\ (bits, bidirectional, 915\ MHz\ @\ 1\ Mbps) = 479{,}257$ bps

$Throughput\ (bits, unidirectional, 2.4\ GHz\ @\ 1\ Mbps) = 817{,}311$ bps

$Throughput\ (bits, bidirectional, 2.4\ GHz\ @\ 1\ Mbps) = 463{,}094$ bps

$Throughput\ (bits, unidirectional, 2.4\ GHz\ @\ 2\ Mbps) = 1{,}308{,}986$ bps

$Throughput\ (bits, bidirectional, 2.4\ GHz\ @\ 2\ Mbps) = 832{,}441$ bps

These calculated metrics are the absolute best case the implemented radio can achieve, given the hardware specific transceiver delays measured.

**4.2.2 Spatial Efficiency**

Given the metrics calculated in section 4.2.1, the spatial efficiency of the CR protocol is given in Table 8. The spatial efficiency of the unidirectional connection is calculated as Throughput / d where d is the modulation data rate. The spatial efficiency of the unidirectional connection is calculated as Throughput / d * 2 where d is the modulation data rate.

**Table 9 – Maximum Theoretical Throughput and Spatial Efficiency**

| Modulation | Throughput | | Spatial Efficiency | |
|---|---|---|---|---|
| | Unidirectional | Bidirectional | Unidirectional | Bidirectional |
| 915 MHz @ 200 Kbps | 181,510 bps | 97,093 bps | 90.7% | 97.1% |
| 915 MHz @ 1 Mbps | 874,794 bps | 479,257 bps | 87.5% | 95.8% |
| 2.4 GHz @ 1 Mbps | 817,311 bps | 463,094 bps | 81.7% | 92.6% |
| 2.4 GHz @ 2 Mbps | 1,308,986 bps | 832,441 bps | 65.4% | 83.2% |

**4.2.3 Parallelization of data transmission and reception**

A unique feature of the CR is simultaneous transmission and reception using multiple transceivers during periods of high data throughput. The CR design implemented can successfully parallelize a serial data stream onto multiple transmitters and on re-serialize the data on the receiver:

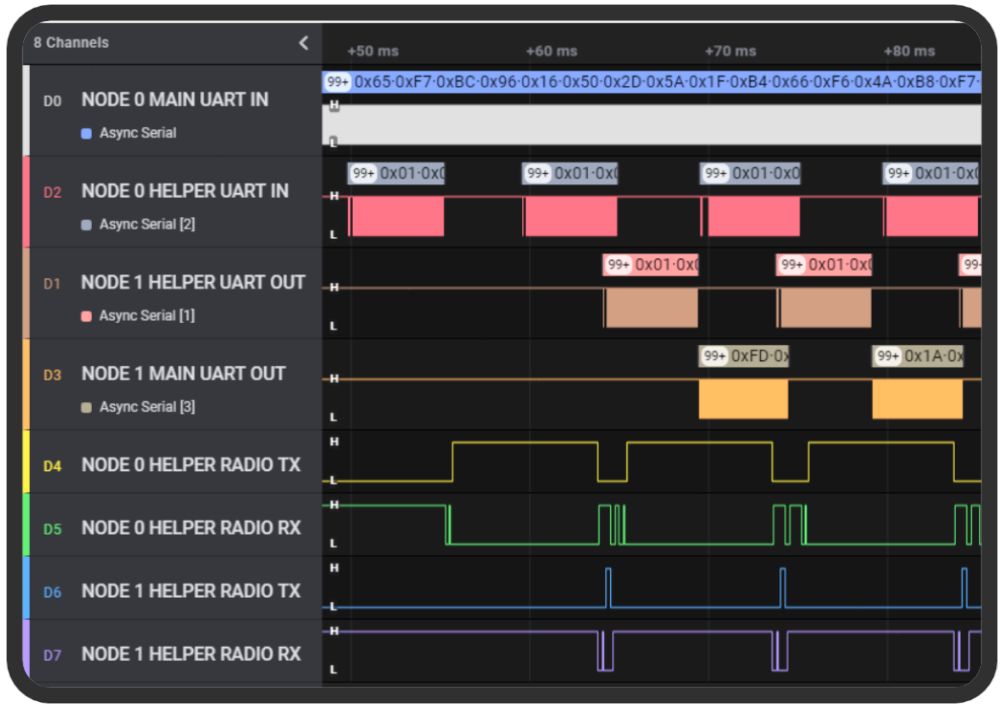

Figure 19 – Helper Data Parallelization

## 4.3 Results

### 4.3.1 Cost Breakdown

The implemented design leverages two Texas Instruments development kits that each contain a CC1352, integrated debugger, various passive components that form the RF frontend, and an integrated voltage regular and high accuracy crystal. The cost breakdown of the implemented CR is shown in Table 10.

Table 10 – Development Kit-Based Hardware CR Cost

| | |
|---|---|
| LAUNCHXL-CC1352P-1 | $ 49.99 |
| LAUNCHXL-CC1352P-2 | $ 49.99 |
| | |
| Total | $ 99.98 |

This cost is not much less than other CRs in this price range, like the Ubertooth One ($150). However, many CRs still only contain one transceiver in their designs, while the implemented CR has two. Additionally, it would easily be possible to cost reduce the development kit-based design greatly with a custom PCB design. Many parts included in the development kit PCB are not used in the CR and would not be necessary in a final real-world design. At volumes of 1000 pieces, such a design is shown in Table 11.

Table 11 – Custom Hardware CR Cost

| Part | Price |
|---|---|
| TI CC1352 | $ 3.90 |
| 2 Layer PCB (1.25" x 1.75") | $ 2.18 |
| SMA Antenna Connector | $ 0.88 |
| Passive Analog SMD Parts (Caps, Resistors, etc.) | $ 2.00 |
| Antenna | $ 2.00 |
| Voltage Regulator | $ 1.87 |
| External High Accuracy Crystal | $ 0.41 |
| | |
| Total | $ 13.24 |
| Total (Main and Helper) | $ 26.48 |

**4.3.2 Latency**

Latency is defined as the amount of time taken for the first byte of a data stream to be accepted as input from the host device, until that byte is outputted to the other node's host device. As in all wireless links, the maximum possible latency is unbounded. In the implemented CR, the minimum latency measured for the CR depends on the modulation data rate, frequency band used, and frame length active. A graph that shows the effect that frame length and modulation data rate have on latency is shown in Figure 20. As can be seen, as the data rate increases, latency deceases. As the frame length increases, latency increases.

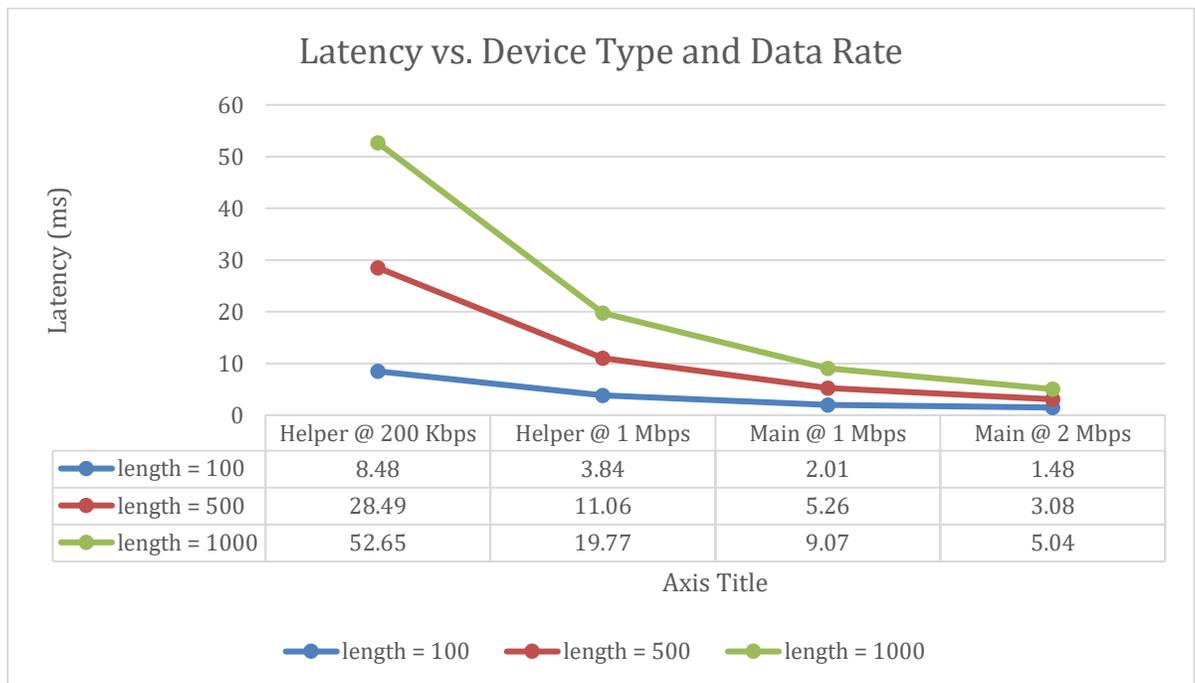

**Figure 20 – Latency vs Device Type and Data Rate**

Helper devices suffer an additional overhead that affects the latency of the frame, due to the fact that data must pass from the initial host device to the sending node's Main, to the sending node's Helper where the frame is transmitted, received by the corresponding receiving node's Helper, back to the receiving node's Main, and then finally outputted to the receiving node's host device. This data flow process is shown in Figure 21.

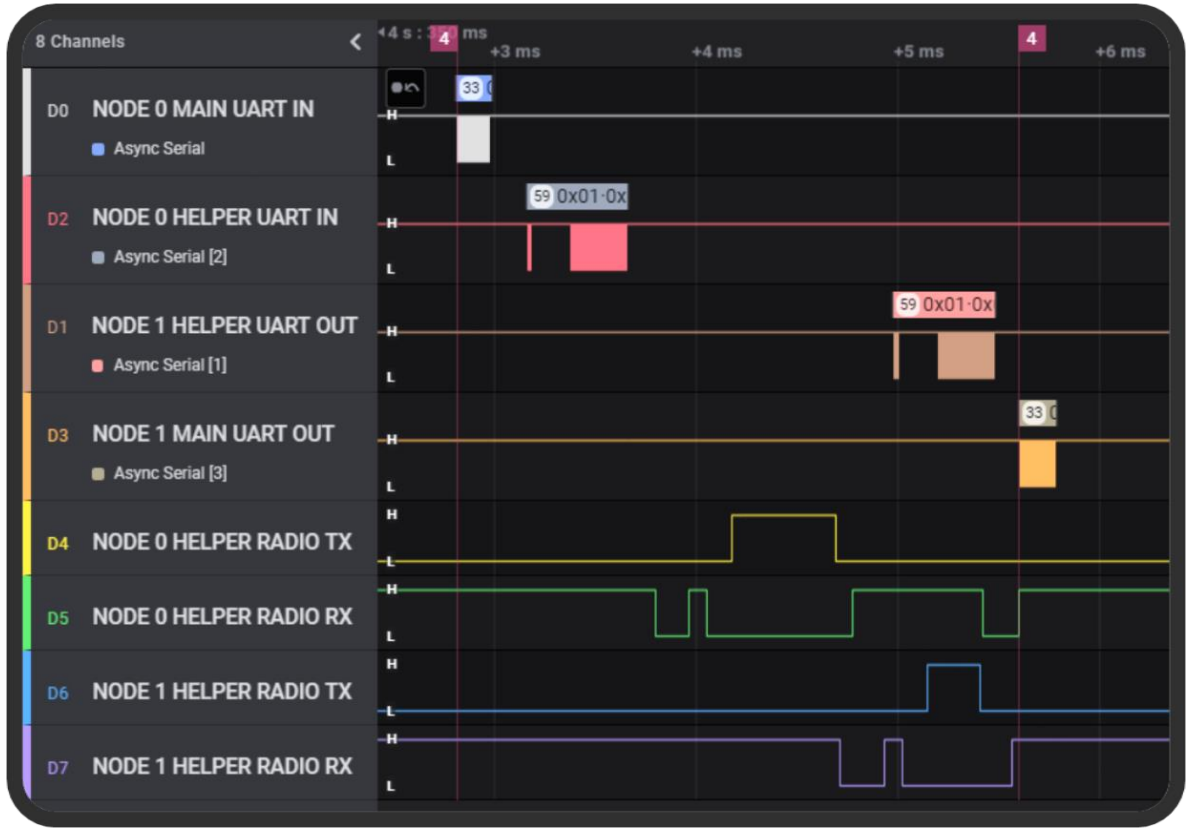

**Figure 21 – Additional Latency Overhead of a Frame Passed through the Helper**

**4.3.3 Range**

The range of the implemented CR depends on many factors, such as the CR's environment, noise floor, possible jammers or inference at the antenna port, antenna gains of the transmitter and receiver, antenna positioning, and whether there is a clear Line of Sight (LOS) between the nodes. Additionally, since the implemented CR has two transceivers that operate on different frequencies, the range is not the same for both the Main and Helper. Lastly, the CR can change its modulation, which also affects the range.

**Table 12 – Friis-Equation Constant Parameters**

| Parameter | Value |
|---|---|
| Antenna locations (height over surface) | 1m |
| Antenna position | Vertical |
| Environment Noise Floor Interference | -125 dBm |
| Transmitter conducted output power | 20 dBm |
| Mean Effective Gain of Antenna | -3 dBi |

By using the Friis-Equation and 2-Ray Ground Reflection Model, with the following above parameters in Table 12, an expected range for each frequency, data rate and modulation are generated:

**Table 13 – Frequency, Data Rate and Modulation vs. Range**

| Frequency, Data Rate and Modulation | Range (GRM) | Range (LOS) |
|---|---|---|
| 915 MHz @ 200 Kbps GFSK | 711 m | 4402 m |
| 915 MHz @ 1 Mbps GFSK | 484 m | 2447 m |
| 2.4 GHz @ 1 Mbps GFSK | 306 m | 918 m |
| 2.4 GHz @ 2 Mbps GFSK | 231 m | 563 m |

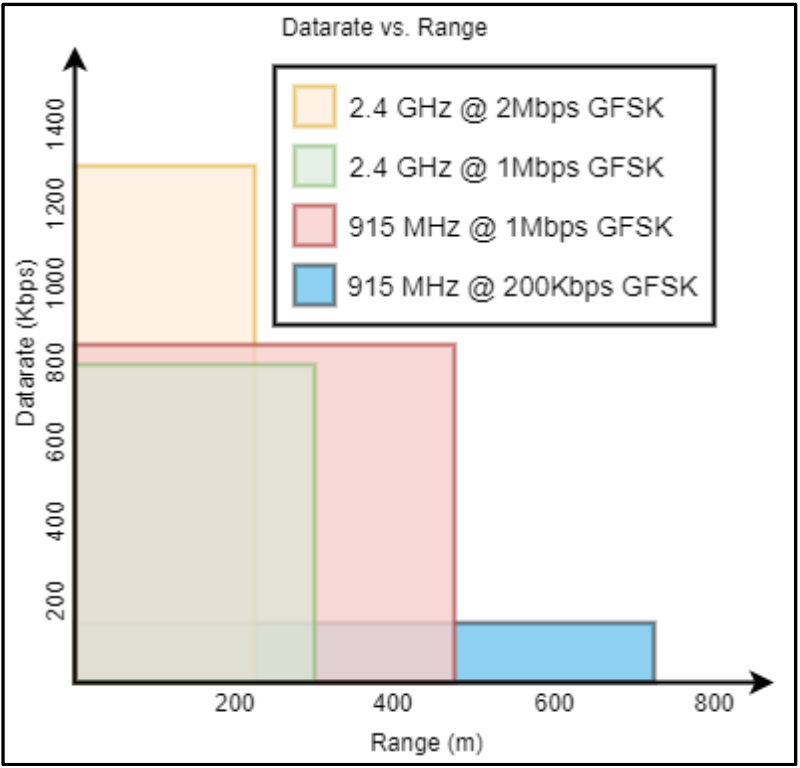

**Figure 22 – Range Overlap Diagram (GRM)**

**4.3.4 Maximum Burst Throughput**

Maximum burst throughput is achieved when the input queue to the CR is entirely full and there is no delay when waiting to send data, and the other node in the CR does not want to send data. To test maximum burst throughput, Test #1 from 4.1 is leveraged to send a frame of 8000 bytes through both the Main and Helper, and the measured are tabulated in Table 14.

**Table 14 – 8000 Byte Burst Throughput**

| Frequency, Data Rate and Modulation | Time Taken for 8000 Bytes | Burst Throughput |
|---|---|---|
| 915 MHz @ 200 Kbps GFSK | 370.61 ms | 172.7 Kbps |
| 915 MHz @ 1 Mbps GFSK | 87.34 ms | 732.8 Kbps |
| 2.4 GHz @ 1 Mbps GFSK | 85.79 ms | 746.0 Kbps |
| 2.4 GHz @ 2 Mbps GFSK | 54.01 ms | 1184 Kbps |

The result for 2 Mbps is slightly lower than would be expected given the computation completed in section 4.2.2. It was found that this is because the Main device cannot keep up at a full 2 Mbps data rate input with the current UART driver utilized.

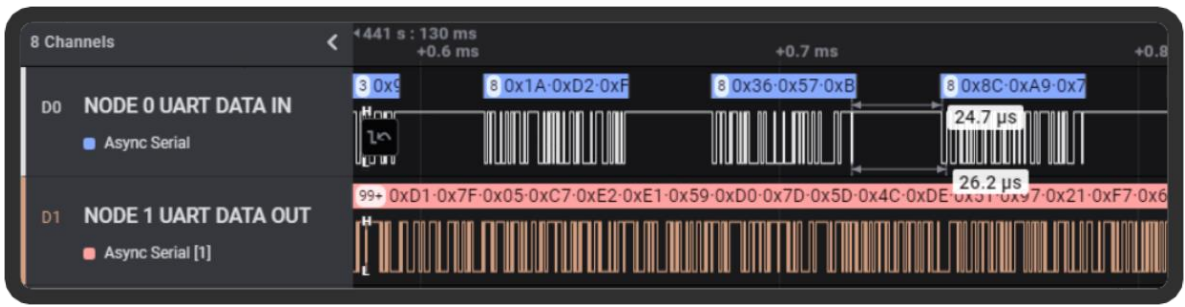

**Figure 23 – Data Rate Bottleneck Due to UART Driver**

The effective data rate measured due the UART driver issue reduces the maximum UART input rate to approximately 1.218 Mbps, which matches closely to our results for the 2 Mbps burst throughput, and shows that it is currently a bottleneck in the implemented CR.  Solutions to this are presented in section 6.

# 5. APPLICATIONS

The CR presented has several advantages over pre-existing RF technologies – it has better congestion resilience due to being able to use non-noisy spectrum, has low latency, and its range and data rate adjusts dynamically with link quality indicators.

## 5.1 Prototype Urban Wi-Fi Uplink

The CR presented could possibly be used in smart-city infrastructure that allows internet connectivity to inner-city users by providing Wi-Fi Access Points through an On-Board Unit (OBU) that receives the wireless link from a Roadside Unit (RSU).

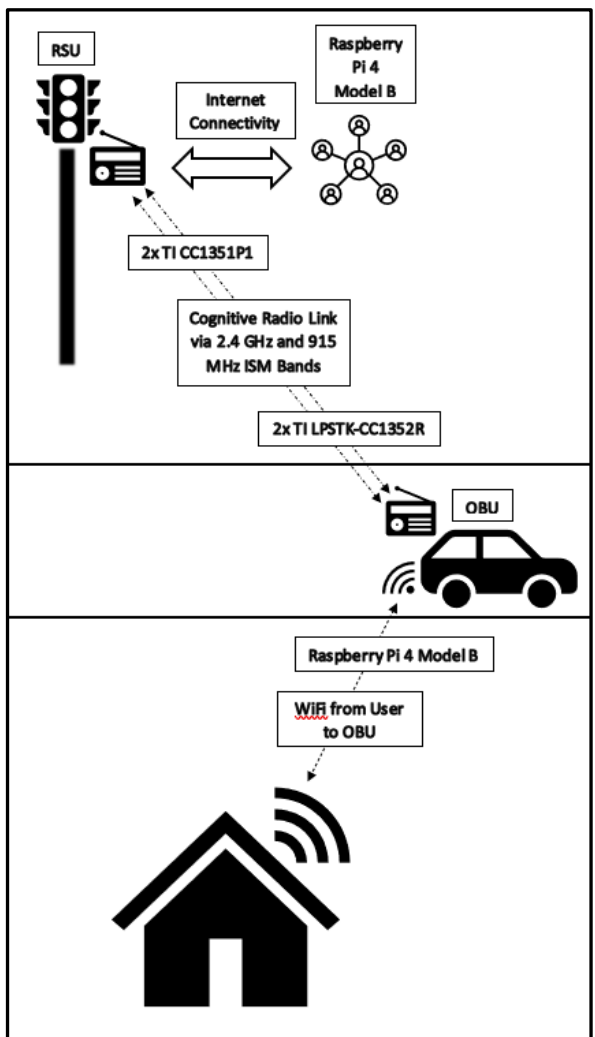

**Figure 24 – Cognitive Radio Uplink Utilized by Smart City Infrastructure**

A prototype of the system above was constructed using two CR nodes and two Raspberry Pi 4 B's. The Raspberry Pi supplying internet connectivity over a Point-to-Point (PPP) data layer link. One Raspberry Pi is connected to ethernet, to provide for an internet uplink, and the other supplies a Wi-Fi access point so that users may connect to the internet through their smartphone or laptop.

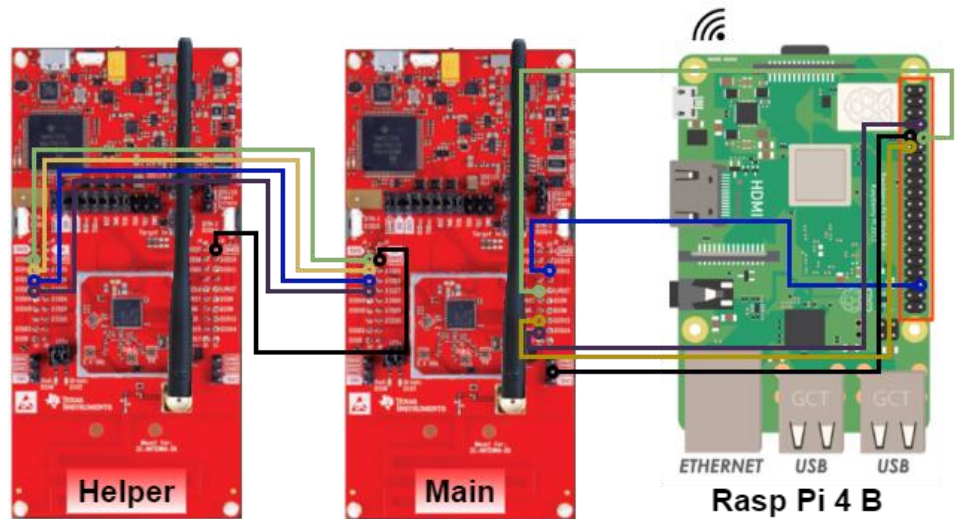

**Figure 25 – Urban Wi-Fi Uplink RSU**

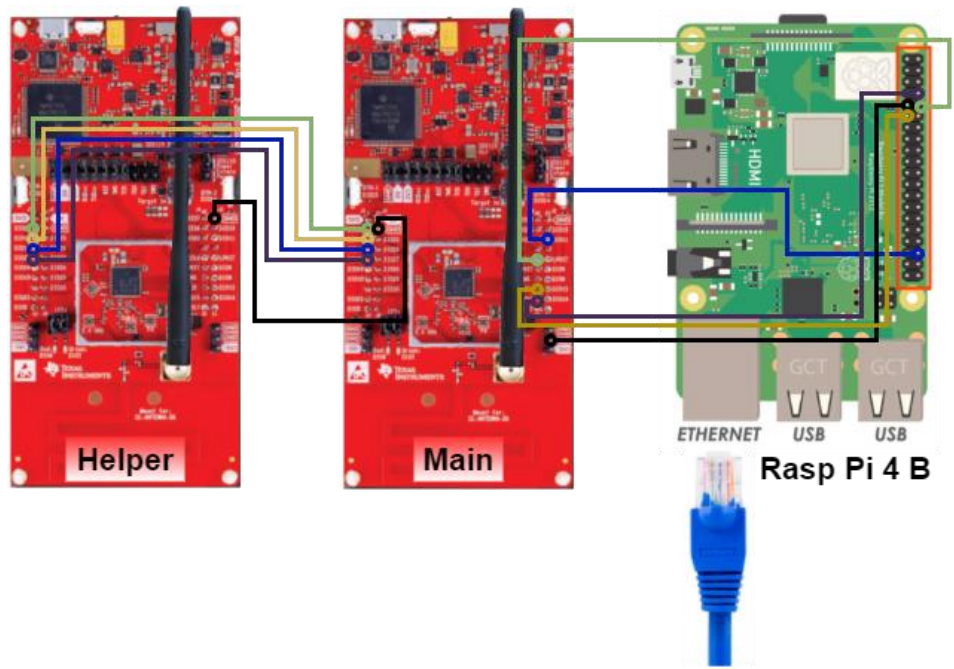

**Figure 26 – Urban Wi-Fi Uplink OBU**

### 5.1.1 Cost

The cost of the parts used in the implemented prototype are outlined in Table 10.

**Table 15 – Urban Wi-Fi Uplink Cost**

| Urban Wifi Uplink as Implemented | Price | # | Cost |
|---|---|---|---|
| LAUNCHXL-CC1352P-1 | $ 49.99 | 2 | $ 99.98 |
| LAUNCHXL-CC1352P-2 | $ 49.99 | 2 | $ 99.98 |
| Raspberry Pi 4 B | $ 55.00 | 2 | $ 110.00 |
| Total | | | $ 199.96 |

However, a cost reduced version is theoretically possible, which cuts the cost of the system by over half. This is only possible by using a cost reduced Custom Hardware CR and the much cheaper Raspberry Pi Zero W.

**Table 16 – Cost Reduced Urban Wi-Fi Uplink Cost**

| Cost Reduced Urban Wi-Fi Uplink | Price | # | Cost |
|---|---|---|---|
| CR Total (Main and Helper) | $ 26.48 | 2 | $ 52.97 |
| Raspberry Pi Zero W | $ 15.00 | 2 | $ 30.00 |
| | | | |
| Total | | | $ 82.97 |

### 5.1.2 Design and Implementation

The Raspberry Pi 4 B acts as the host device to both nodes in the CR and connects over the UART interface running at 2 Mbps. To configure the Raspberry Pi 4 B to use a dedicated UART interface capable of speeds up to 2 Mbps, with hardware flow control, instead of a software-emulated one only capable of slower speeds and no flow control, Bluetooth hardware must be turned off as this uses the dedicated UART. This was done through modification of the device trees in the Raspbian kernel configuration.

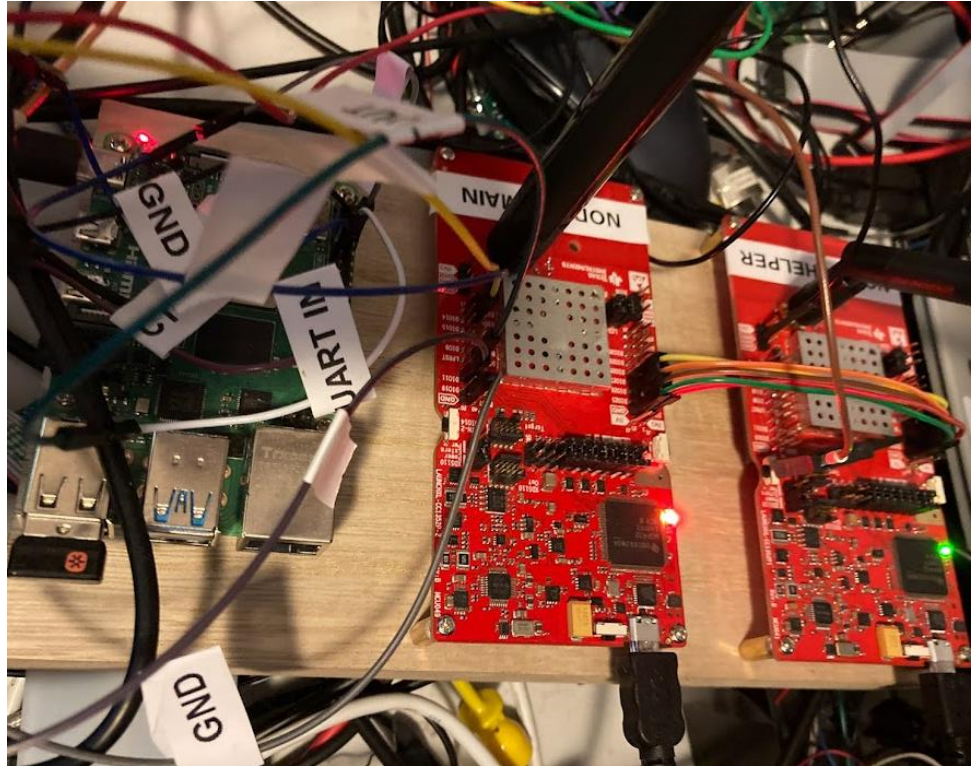
**Figure 27 – Prototype Urban Wi-Fi Uplink Node**

The CR wireless link is transparent to the Raspberry Pi connection, which is made over a PPP link. The OBU node in the system provides a Wi-Fi Access Point to the user and is created by using the RaspAP utility. The RaspAP utility can take the PPP link that the CR provides from the RSU which provides internet connectivity via its wired ethernet port.

## 5.1.3 Prototype End-to-End Performance (Wi-Fi Speed Test)

End to end performance tests were performed with the created prototype. At the OBU, the Wi-Fi hotspot is created using another Raspberry Pi. Both nodes are stationary. From there, a connection is made as a Wi-Fi station on an iPhone SE 2020. Various speed tests were performed with the CR link in three states.

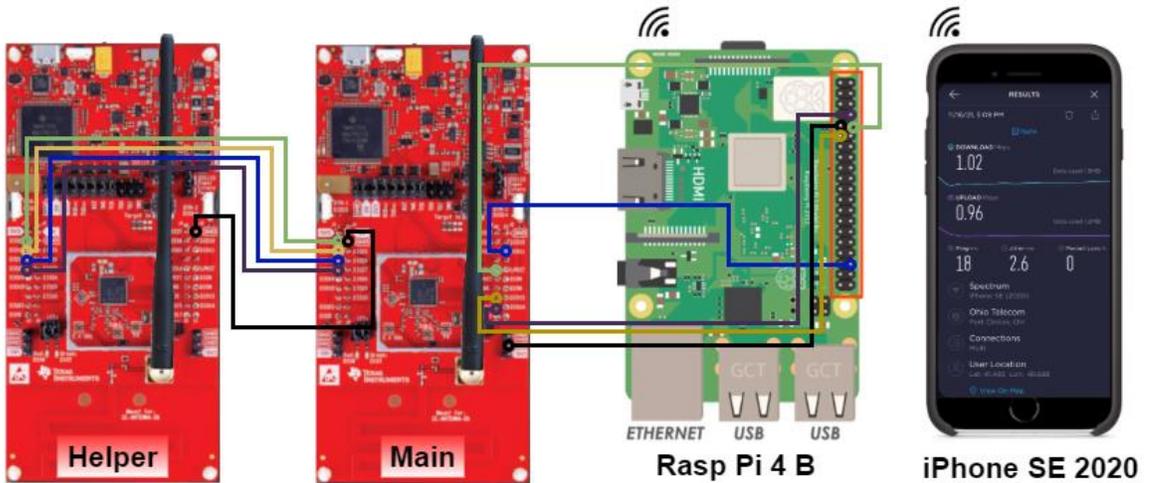
Figure 28 – Wi-Fi Speed Test Setup

The first test completely blocks the Helper from being able to send frames.  The speed test results are shown in the figure below.

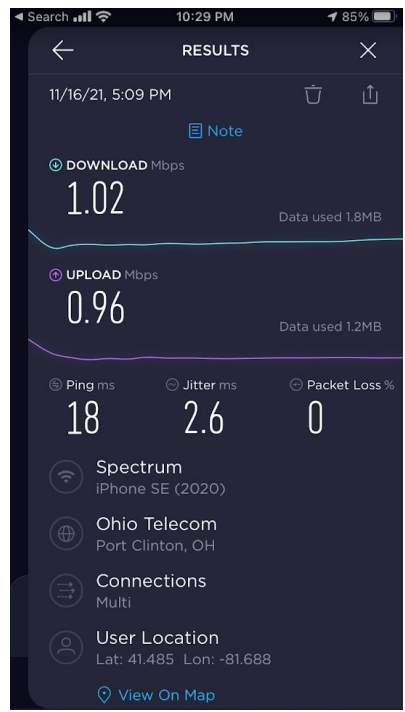
Figure 29 – Helper Blocked Speed Test

The second test completely blocks the Main from being able to send frames.  The speed test results are shown in the figure below.

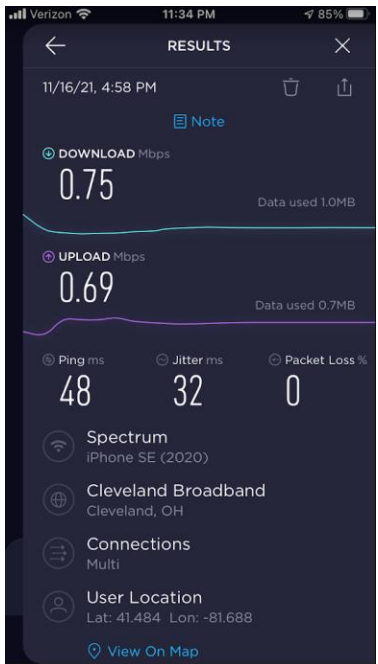
Figure 30 – Main Blocked Speed Test

The third test allows both the Main and Helper to send frames.  The speed test results are shown in the figure below.

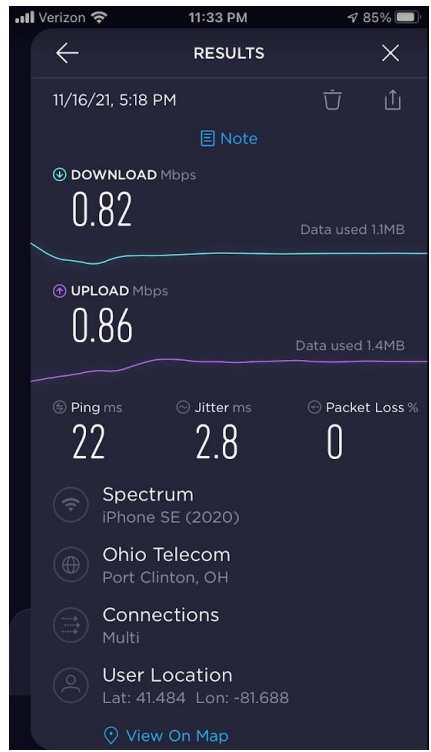

Figure 31 – CR Node Full Functionality Speed Test

The third test shows a slight decrease in performance, but this is due to the same UART driver problem described in Section 4.3.4. Workarounds are proposed in Section 6.

## 5.2 Low-Cost Drone Communication

Drones are another application well suited to the proposed protocol since the implemented CR is resilient to congestion by using multiple frequency bands simultaneously. Already, the drone industry has been to adopt Cognitive Radios for consumer use. For example, DJI has implemented an RF protocol called Occusync 2.0, which is a propriety RF protocol that can use either 2.4 GHz or 5.8 GHz band frequencies. The implemented protocol has an advantage over Occusync 2.0 in that is operates on the 915 MHz band, which allows for improved penetration and range. Multi-band CR communication could be well suited for drones, such that the critical command and control signals use band that is incurring the least interference, while the less critical functionalities like video feeds and other telemetry use the secondary bands.

## 5.3 DSRC

Dedicated Short-Range Communications (DSRC) technology describes one-way or bi-directional wireless communication protocols specifically designed for automotive use. The US Department of Transportation purports that DSRC technologies will be leveraged to enhance safety and improve performance of cooperative automated driving systems, technology that has only just begun to be developed. To that end, in 1999 the FCC allocated 75 MHz of spectrum in the 5.9 GHz band for DSRC-only applications. However, citing lack of widespread adoption, the FCC recently reallocated the entirety of DSRC's spectrum for Wi-Fi and C-V2X applications. Given the FCC's recent November 2020 decision to remove licensed-band spectrum specifically allocated for DSRC, the CR presented could benefit DSRC technology since it provides extra reliability by using multiple transceivers simultaneously. Additionally, the CR design presented in this thesis seeks to circumvent common barriers to widespread industry adoption, such as cost and complexity by using off-the-shelf components.

# 6. FUTURE WORK

The implemented design only contains two transceivers, and in theory could be extended to any number of transceivers. However, a different interface to connect the transceivers together would be necessary, as daisy chaining Helper UARTs to pass a frame from the Main to a Helper at the end of the chain could be very inefficient. To add more transceivers on the CC1352, it would be prudent to use the multiple SPI interfaces, as well as utilize the SPI's Chip Select (CS) functionality in order to select which device is receiving the data from the main.

In this thesis, the CR's protocol is merely a point-to-point wireless connection. It would be desirable to extend this protocol to multi-point, such that there could be more than 2 nodes in the system. By using an address field in the frame, packets could be shuffled along or filtered when received at their destination.

The CR's Radio protocol does not attempt any channel switching or frequency hopping. In theory, the CR could add some method to channel hop. If a channel is blocked for too long, the nodes in the system could hop to a previously determined channel.

A major bottleneck in the implemented design is the UART interface to the host device, due to the fact that the UART hardware cannot accept data at faster than 1.3Mbps during peak data transfer. This should be examined, and it is suspected that utilizing the Direct Memory Access (DMA) features of the CC1352 may allow the bus to accept data at a faster speed. This suspicion is due to the fact upon inspection, it was revealed that the current driver is copying each data byte by byte from a circular queue. This means that CPU is involved with each byte received or sent. However, DMA could be used to free the CPU from the hardware peripheral, such that instead of the CPU touching each byte, it only needs to point the UART hardware to an address in memory and begin the transfer. Additionally, there are some instances in the implemented software design where zero-length copies could replace full copies of frames into queues, such as from the Sequencer to the Main's RF Control task.

Lastly, a higher speed PHY could be used. Although the CC1352 does not allow for modulation data rates above 2Mbps (and 1Mbps in the 915 MHz band), other related chipsets allow for up to 5Mbps. However, as described in Section 4.2.3, this would result in a range trade off. To truly benefit from a faster on-air data rate, the data throughput bottleneck from the UART input stream would need to be addressed first.

# 7. CONCLUSION

This thesis presents a low-cost hardware platform built from off-the-shelf components that utilizes free to use Industrial, Scientific, and Medical (ISM) bands, and implements a concurrent multi-spectrum point-to-point wireless protocol optimized for non-stationary devices. Additionally, the software architecture necessary to program the logic needed to parallelize and re-sequence the frames sent on multiple transceivers in parallel in a performant method is described and implemented. Performance metrics such as cost, latency, throughput, and range are measured and analyzed. Applications of such a wireless implementation are proposed and implemented, such as smart-city infrastructure that allows internet connectivity to inner-city users by providing Wi-Fi Access Points through mobile On-Board Unit (OBU) devices with uplinks delivered from stationary Roadside Unit (RSU) devices.